\documentclass[preprint,12pt]{elsarticle}
\biboptions{sort&compress,longnamesfirst,numbers}
\usepackage[english]{babel}
\usepackage{geometry}
\usepackage{pdflscape}
\usepackage{lineno}
\usepackage{multirow}
\usepackage{graphicx}
\usepackage{tabularx}
\usepackage{mathtools}
\usepackage{threeparttable}
 \newcolumntype{L}{>{\raggedright\arraybackslash}X}
\usepackage{booktabs}
\usepackage{soul}
\usepackage[export]{adjustbox}
\usepackage{subfig}
\usepackage{amsmath}
\usepackage{amssymb}
\usepackage[titletoc]{appendix}
\usepackage{siunitx} 
\usepackage{colortbl}
\usepackage{url}

\usepackage{breakurl}
\usepackage[breaklinks]{hyperref}
\definecolor{kugray5}{RGB}{224,224,224}
\makeatletter 
\renewcommand\@biblabel[1]{#1.} 
\makeatother
\usepackage{floatrow}
\usepackage[justification=centering]{caption}
\begin{document}
\begin{frontmatter}
\title{Techno-Economic Case Study of a Rural Local Electricity Community in Switzerland}
\author[EPFL]{Gerard Marias Gonzalez\corref{cor1}}
\author[EPFL]{Alejandro Pena-Bello}
\author[EPFL]{Jérémy Dumoulin}
\author[EPFL]{Nicolas Wyrsch}
\cortext[cor1]{Corresponding author. E-mail: gerard.mariasgonzalez@epfl.ch}

\address[EPFL]{Photovoltaics and thin film electronics laboratory (PV-LAB), École Polytechnique Fédérale de Lausanne (EPFL), Institute of Electrical and Microengineering (IEM), Neuchâtel, Switzerland}

\begin{abstract}
Local Electricity Communities (communautés électriques locales, CEL) will become operational in Switzerland in 2026, allowing prosumers, consumers, and storage operators within the same municipality and distribution system operator (DSO) area to exchange electricity over the public grid with reduced distribution tariffs. This report examines a rural Swiss case study to explore the techno-economic implications of CELs for both participants and the local DSO. The findings indicate that CELs can enhance the local use of renewable generation, particularly photovoltaics, and offer modest financial gains, with outcomes strongly shaped by community size, composition, and tariff design. Larger and more heterogeneous communities achieve better internal matching of supply and demand, though the overall incentive remains limited because the tariff reduction applies only to distribution charges. The study further shows that internal energy exchange is maximized when local PV generation covers roughly 1–2 times the community load. For DSOs, CELs reduce grid imports (27–46\%), resulting in a substantial reduction in distribution tariff revenues (17–36\%), necessitating regulatory adaptation. While centralized batteries provide economic value to members, their technical impact on the grid remains modest due to their small, economically optimized capacity. Larger centralized storage is shown to reduce transformer peak power, but risks increasing line loading, suggesting a need for careful sizing and placement.

\end{abstract}
\end{frontmatter}
\section{Introduction}
Local Electricity Communities (communautés électriques locales, CEL) are Swiss arrangements that let prosumers, consumers and storage operators within a single municipality and distribution system operator (DSO) service area to exchange locally generated electricity over the public distribution grid \cite{Switzerland_Fedlex_2024_OC679}; participation requires smart metering, the community must embed a minimum share of renewable generation equal to 5\% of the sum of participants’ connection capacities, exchanges that are simultaneous and occur within the CEL benefit from a reduction of the network usage tariff (40\% when all members are connected to the same low-voltage network, or 20\% when they are on different low-voltage networks behind the same transformer), and only distribution levels 5 and 7 (i.e.\,$\leq$ 36 kV) are permitted. These rules are scheduled to take effect on 1 January 2026 \cite{Switzerland_Fedlex_2024_OC679}.  

CELs differ fundamentally from self-consumption clusters (i.e. RCP in french or ZEV in german) and their virtual form (RCPv and vZEV, respectively): an RCP aggregates several end-users behind a private sub-network and acts as a single customer with respect to the DSO, while an RCPv uses the DSO’s meters and connection lines to virtually sum dispersed buildings into one billing entity; neither scheme enables market trading among distinct customers over the public network. By contrast, CELs allow producers and consumers within the same municipality and connected to the same DSO to exchange locally produced renewable electricity over the public LV/MV grid, while each participant keeps an individual supply contract with the DSO; internal exchanges benefit from a reduced network-usage tariff when no transformer is crossed \cite{LApEl_OApEl_Art,Debons_2025_MiniLiberationMarcheElectricite}. The operator or representative of a CEL must interact administratively with the DSO and is responsible for settlement, billing, and internal management of the community. CELs thus introduce a “mini-liberalization” at the municipal scale, allowing local energy trading and contractual innovation while maintaining technical coordination by the DSO \cite{Debons_2025_MiniLiberationMarcheElectricite}.

Participation in a CEL is not mandatory; consumers and producers remain free to join or not, which allows multiple contractual and tariff arrangements to coexist in the same area. Smart meters are required for all CEL participants, and the local DSO must provide granular,  automated energy flow data in the Swiss SDAT standard \cite{VSE_SDAT_CH_2022}, ensuring the internal allocation of costs and benefits is managed transparently and efficiently. Each site (prosumer, consumer, or storage) may only belong to one CEL at a time, ensuring clear boundaries and governance within the energy community model. The operator or representative of a CEL acts as the administrative interface with the DSO and is responsible for tasks such as internal billing, tracking member participation, and energy settlement. 

In developed economies, energy communities are increasingly viewed as innovative solutions to boost energy self-sufficiency, facilitate local use of renewables (especially PV), and support climate targets by reducing reliance on fossil fuels  \cite{Bellini_Campana_Censi_DiRenzo_Tarola_2024}. Urban communities have dominated the literature, while rural energy communities, despite their high technical potential and unique opportunities and challenges, have received far less attention. Notably, studies consistently highlight the strong potential of rural energy communities as a driver for regional development and grid resilience \cite{Doloi_Crawford_Varghese_2022_SmartVillagesRural}.

In Switzerland, Romano and Trutnevyte recently found that rural CELs could generate 8 TWh  annually out of the targeted 34 TWh for 2050 \cite{OFEN2050}, or 23\% of Switzerland’s 2035 renewable target and 43\% of rural demand \cite{Romano_Trutnevyte_2025_LEC_Rural_Switzerland}. However, the actual output may fall to 4 TWh if investment strategies prioritize financial returns over social benefits. Schnidrig et al.  \cite{Schnidrig2024Districts} showed that although CELs provide significant technical and economic advantages over traditional models, such as greater revenue retention and fairer cost and benefit distribution, a successful transition requires utilities and DSOs to adapt their business models, particularly through tailored local distribution tariffs. These adaptations are essential for balancing stakeholder interests and supporting the broader adoption of decentralized, community-based energy systems.

From a regulatory perspective, CELs constitute a partial market liberalization at the municipal scale: they enable bilaterally negotiated renewable electricity supply contracts between local actors, undermining centralized supply monopolies but retaining technical coordination via the DSO. Notably, the network usage tariff reduction, which is typically 40\% for exchanges strictly within the low-voltage grid and decreases to 20\% if the community utilizes the medium-voltage network, applies only to distribution charges. It does not apply to energy taxes or other fees, meaning the net financial incentive for CEL participation may remain limited.

This report aims to assess the techno-economic benefits of a rural CEL by jointly considering participant benefits and DSO implications across different tariff schemes, sizes, and configurations. The rural case study was selected with the expectation that its spread network structure would be more sensitive to local generation and consumption dynamics, providing a suitable context to explore the potential technical and economic impacts of CEL formation.

\section{Methodology and Data}
\subsection{Demand and PV data}
Due to a lack of public availability of building-level demand measurements in low-voltage networks, we infer each building’s 15-minute load curve by combining three datasets, following the approach described in~\cite{penabello2024arxiv}. First, we use static attributes from the \textit{Registre fédéral des bâtiments et des logements} (Federal Register of Buildings and Dwellings, RegBL)~\cite{RegBL}, year of construction/renovation, use category, conditioned floor area, and number of storeys, characterize the stock. Second, we rely on Swiss SIA standards for electricity-intensity benchmarks by building type and vintage. Third, we use anonymized 15-minute smart-meter load curves from the Flexi projects in French-speaking Switzerland~\cite{Flexi1_2015, Flexi2_2017, holweger2024privacy}, which provide representative profiles for apartments, single-family houses, and non-residential buildings. These proxy load shapes are then assigned and scaled to each building based on its floor area (from RegBL~\cite{RegBL}) and its corresponding SIA intensity, before being reconciled to ensure that the aggregated demand matches the transformer-level measurements supplied by Romande Energie.

Roof geometry (usable area, azimuth, tilt) for each building in the studied LV grid is extracted from the RegBL. Photovoltaic generation is simulated using monitored outdoor air temperature and global horizontal irradiance from the MeteoSwiss station in Pully (Vaud). Module electrical parameters are taken from the Sandia PV database entry for the SunPower SPR-315E-WHT (monocrystalline, 19.3\% efficiency, module area $1.6310~\mathrm{m}^2$) and combined with the meteorological series to compute time-step current–voltage and power per module. Arrays are assembled per roof plane and aggregated by orientation to obtain building-level PV output. The maximum feasible capacity per building corresponds to 70\% of the building’s available rooftop area.

For the 2025 reference year, PV and battery capital costs are taken from IRENA and calibrated to Swiss price levels using the SuisseÉnergie market study via the empirical adjustment of Bloch et al.  \cite{irena_esr2017,suisseenergie2021,bloch2019impact}. Investments are represented by an affine cost function, 

\begin{equation}    
C(x)=C_0+c_1 x
\end{equation}

where \(C_0\) covers permitting, design, and other fixed installation items, and \(c_1\) scales with installed capacity. These costs are annualized (capital-recovery) and combined with PV annual O\&M and battery lifetime considerations in the objective function; the techno-economic inputs used are summarized in Table~\ref{tab:param}.

\begin{table}[H]
\begin{tabular}{lll}
\toprule
Parameter              & Value  &    Reference  \\\midrule
System lifetime        & 25 years  &   \cite{irena_esr2017} \\
Discount rate          & 3\%        & Own assumption\\
PV fixed cost          & 10 049 \,CHF &   \cite{irena_esr2017,suisseenergie2021}\\
PV specific costs      & 1.05 \,CHF/W &   \cite{irena_esr2017,suisseenergie2021}\\
Battery fixed cost     & 0 \,CHF      &   \cite{irena_esr2017,suisseenergie2021}\\
Battery specific costs & 229 \,CHF/kWh &  \cite{irena_esr2017,suisseenergie2021}\\\bottomrule
\end{tabular}
\caption{Techno-economic parameters used in this study.}
\label{tab:param}

\end{table}

\subsection{Optimization model}
\label{subsection:optimization_model}
The operation of each building and community configuration is represented through an optimization model. 
The objective is to determine the techno-economic configuration that minimizes the total cost of ownership (TOTEX) over the project lifetime. 
TOTEX accounts for two components: operating expenditures (OPEX), which cover grid exchanges, PV maintenance, and battery and inverter replacement, and annualized capital expenditures (CAPEX), corresponding to PV and battery investments.

Formally, the optimization problem is expressed as:
\begin{equation}
    TOTEX = OPEX + R \cdot CAPEX
    \label{eq:obj}
\end{equation}
where \(R\) is the annuity factor that accounts for the system lifetime \(L\) and discount rate \(r\):  
\begin{equation}
    R = \frac{r \cdot (1+r)^L}{(1+r)^L - 1}
\end{equation}

OPEX includes three components: grid exchanges, battery operation, and PV maintenance:  
\begin{equation}
\label{eq:opex}
    OPEX = OX_{ge} + OX_{bo} + OX_{pm}
\end{equation}

\begin{table}[H]
\centering
\begin{tabular}{lll}
\toprule
Name               & Notation & Definition \\\midrule
Grid exchanges     & \(OX_{ge}\) & Costs of imports/exports (Eq.~\ref{eq:ge_vol}) \\
Battery operation  & \(OX_{bo}\) & \( \sum_{t=1}^T (P_{t}^{DIS} \cdot C^{BAT}_d + P_{t}^{CH} \cdot C^{BAT}_c ) \cdot TS_t \) \\
PV maintenance     & \(OX_{pm}\) & \( \gamma^M \cdot CX_{pv} \) \\\bottomrule
\end{tabular}
\caption{Components of OPEX.}
\label{tab:opex_parts}
\end{table}

In these definitions, \(\gamma^M\) denotes the annual specific maintenance cost (expressed as a fraction of the PV investment). The grid exchange costs are calculated as:
\begin{equation}
    OX_{ge}^{vol} = \sum_{t=1}^T \left[ P_t^{IMP} \cdot t_t^{IMP} - P_t^{EXP} \cdot t_t^{EXP} \right] \cdot TS_t
    \label{eq:ge_vol}
\end{equation}
where \(P_t^{IMP}\) and \(P_t^{EXP}\) denote the imported and exported power at time step \(t\), and \(TS_t\) is the time-step duration.  

Battery-related operating costs explicitly account for battery degradation and replacement over the project lifetime. Battery aging is modeled as a function of the state-of-charge (SoC) trajectory, cycling behavior, and elapsed time. A rain-flow cycle-counting algorithm is applied to the SoC profile to identify equivalent charge–discharge cycles with varying depths and mean SoC levels, while accounting for both cycling-induced degradation and calendar aging effects. The loss of battery capacity accumulates over time, and it is assumed that the battery reaches its end-of-life when the remaining usable capacity falls below 80\% of its nominal capacity. When this condition is met, the battery is replaced, and the corresponding replacement cost is included in OPEX. If the final battery replacement occurs before the end of the project lifetime, a residual value proportional to the remaining battery lifetime is credited at the end of the analysis period.

In addition, PV inverter replacement costs are included in OPEX by assuming a fixed inverter lifetime of 15 years. Replacements occur at regular intervals over the project horizon, and a residual value proportional to the remaining inverter lifetime is assigned at the end of the analysis period. Finally, PV maintenance costs are modeled as annual operating expenditures proportional to the initial PV investment, accounting for routine inspections, cleaning, and minor component replacements throughout the system's lifetime.

CAPEX captures the investment costs of PV and battery systems:  
\begin{equation}
\label{eq:capex}
    CAPEX = CX_{pv} + \frac{L}{L^{bat}} \cdot CX_{bat}
\end{equation}

\begin{table}[H]
\centering
\begin{tabular}{ll}
\toprule
Component & Definition \\\midrule
PV        & \( CX_{pv} = \sum_{i=1}^{N} \mu_i^{MOD} \cdot P_{nom,i}^{MOD} \cdot C^{MOD} + \beta^W \cdot C^{FW} \) \\
Battery   & \( CX_{bat} = \sum_{j=1}^{B} E_{cap,j}^{BAT} \cdot C^{BAT}_{sp} + \beta^{BAT} \cdot C^{BAT}_{fix} \) \\\bottomrule
\end{tabular}
\caption{Components of CAPEX for PV and battery.}
\label{tab:capex_parts}
\end{table}

A complete list of parameters used in the optimization model is provided in~\ref{annex:parameters}.

\subsection{Electricity tariffs}
In this study, the applicable tariff structure depends on the origin of the energy. Energy imported from the external grid (DSO) is billed under the standard double tariff (DT). In contrast, internal exchanges within the CEL are evaluated under two distinct pricing models. The first follows the double tariff structure, incorporating a 40\% reduction on distribution grid charges, while the second applies a dynamic, irradiance-based tariff developed as part of this work. Both structures use the 2025 Romande Energie tariff data (including cantonal but excluding communal fees), with the feed-in tariff (remuneration rate) fixed at 11.5 cts/kWh.

For the DT-based approach, internal exchanges follow the same peak and off-peak periods as the external grid tariff. However, a 40\% reduction is applied exclusively to the distribution grid usage component. It is important to note that this discount does not apply to the energy supply costs or public taxes, which remain identical to the standard external tariff. Peak hours are defined from 17:00 to 22:00, Monday to Friday, while all other hours are considered off-peak. Under this scheme, energy imported from the DSO is billed at 36.49 cts/kWh (peak) and 24.87 cts/kWh (off-peak), whereas internal exchanges are billed at 29.83 cts/kWh (peak) and 20.91 cts/kWh (off-peak). A detailed breakdown of the tariff composition is provided in ~\ref{annex:tariff_breakdown}.

\begin{table}[h]
\centering
\caption{Overview of the 2025 tariff structures considered in this study for internal and external exchanges.}
\label{tab:tariffs}
\small 
\begin{tabular}{lc}
\hline
\textbf{Tariff Structure} & \textbf{Price [cts./kWh]} \\
\hline
\multicolumn{2}{l}{\textit{\textbf{External Exchanges (DSO)}}} \\
Double Tariff -- Peak (17:00--22:00) & 36.49 \\
Double Tariff -- Off-peak & 24.87 \\
\hline
\multicolumn{2}{l}{\textit{\textbf{Internal Exchanges (CEL)}}} \\
Double Tariff -- Peak (17:00--22:00) & 29.83 \\
Double Tariff -- Off-peak & 20.91 \\
Dynamic Tariff -- Max Price (0\% Irradiance) & 24.52 \\
Dynamic Tariff -- Min Price (100\% Irradiance) & 11.50 \\
\hline
\end{tabular}
\end{table}

Alternatively, internal exchanges are priced using a dynamic, irradiance-based tariff proposed in this study. The price is inversely proportional to the 15-minute solar irradiance: high irradiance leads to low prices, encouraging the use of locally generated solar energy when it is abundant. It is important to note that the peak/off-peak timing of the external double tariff (17:00-22:00) generally does not align with the maximum/minimum pricing periods of the dynamic tariff. The tariff operates within defined bounds. At zero irradiance, the price is capped at 24.52 cts/kWh, equivalent to the Simple Tariff with the 40\% reduction on the distribution component. At full irradiance, the price reaches its minimum at the feed-in tariff (11.5 cts/kWh). These limits ensure that CEL members always have an incentive to exchange energy. The dynamic tariff is decomposed into fixed taxes, a reduced fixed grid component, and a variable energy component determined as the residual value. The full decomposition procedure is detailed in ~\ref{annex:tariff_breakdown}.

\begin{figure}[H]
\centering
\includegraphics[width=0.8\textwidth]{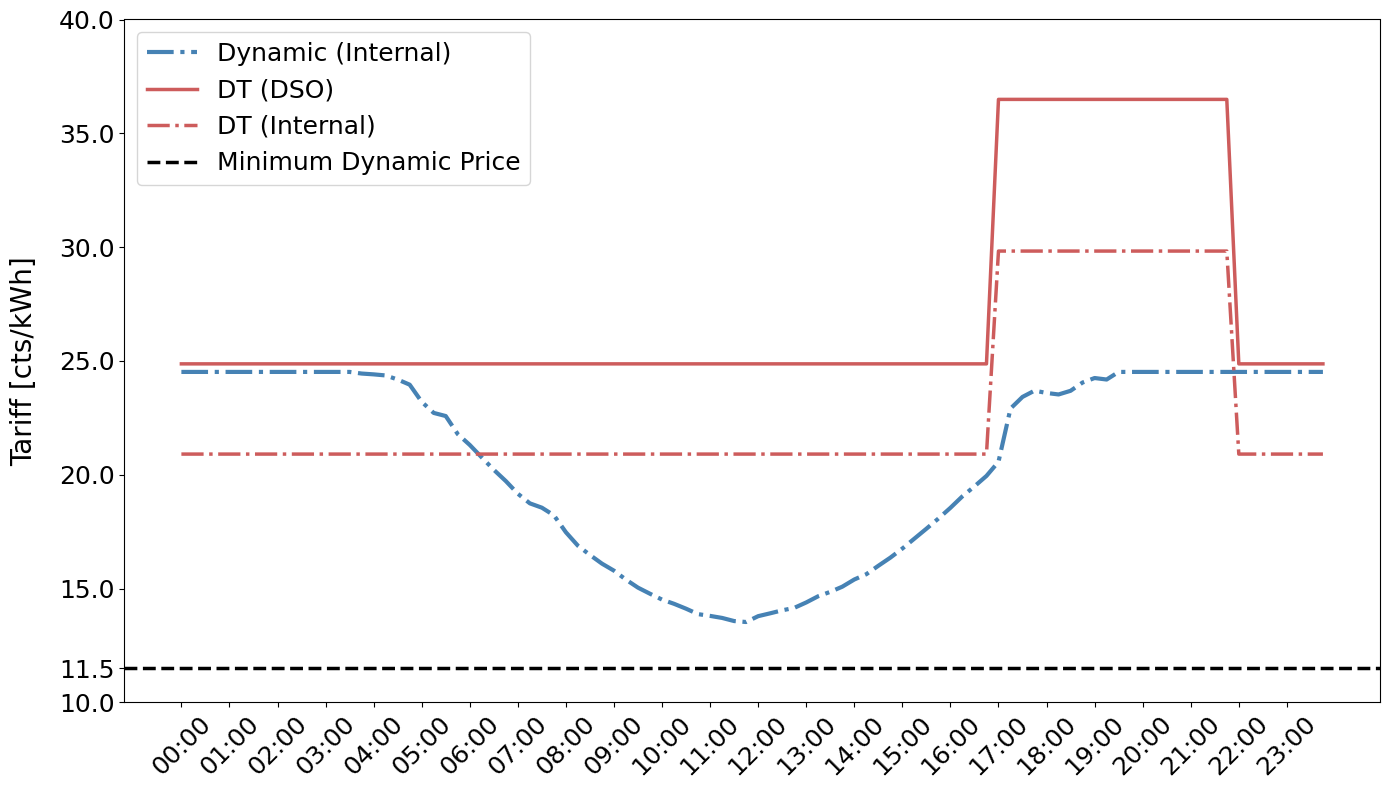}
\caption{Hourly evolution of the tariff schemes using 2025 rates applied to a representative summer day (24~June~2024). The graph compares the external price (DSO) against the internal pricing options (Double and Dynamic), and indicates the minimum dynamic price at 11.5~cts/kWh (dashed line).}
\label{fig:tariff_day}
\end{figure}

Figure~\ref{fig:tariff_day} illustrates the evolution of the pricing schemes throughout a representative summer day (24 June 2024). The graph contrasts the Standard Double Tariff (applied to external imports) against the two available options for internal exchanges: the Internal Double Tariff and the Internal Dynamic Tariff. Both internal tariffs benefit from the 40\% regulatory reduction on distribution fees. Note that this figure uses the 2025 tariffs applied to the 2024 irradiance data (used here to represent the current year, as the 2025 records are not yet available). The black dashed line marks the absolute minimum price floor (11.5~cts/kWh). On this specific day, solar irradiance was not high enough to trigger this floor, so the dynamic tariff remained above this limit throughout the day.

\subsection{Low-voltage network data}
\begin{figure}[H]
    \centering
    \includegraphics[width=0.6\textwidth]{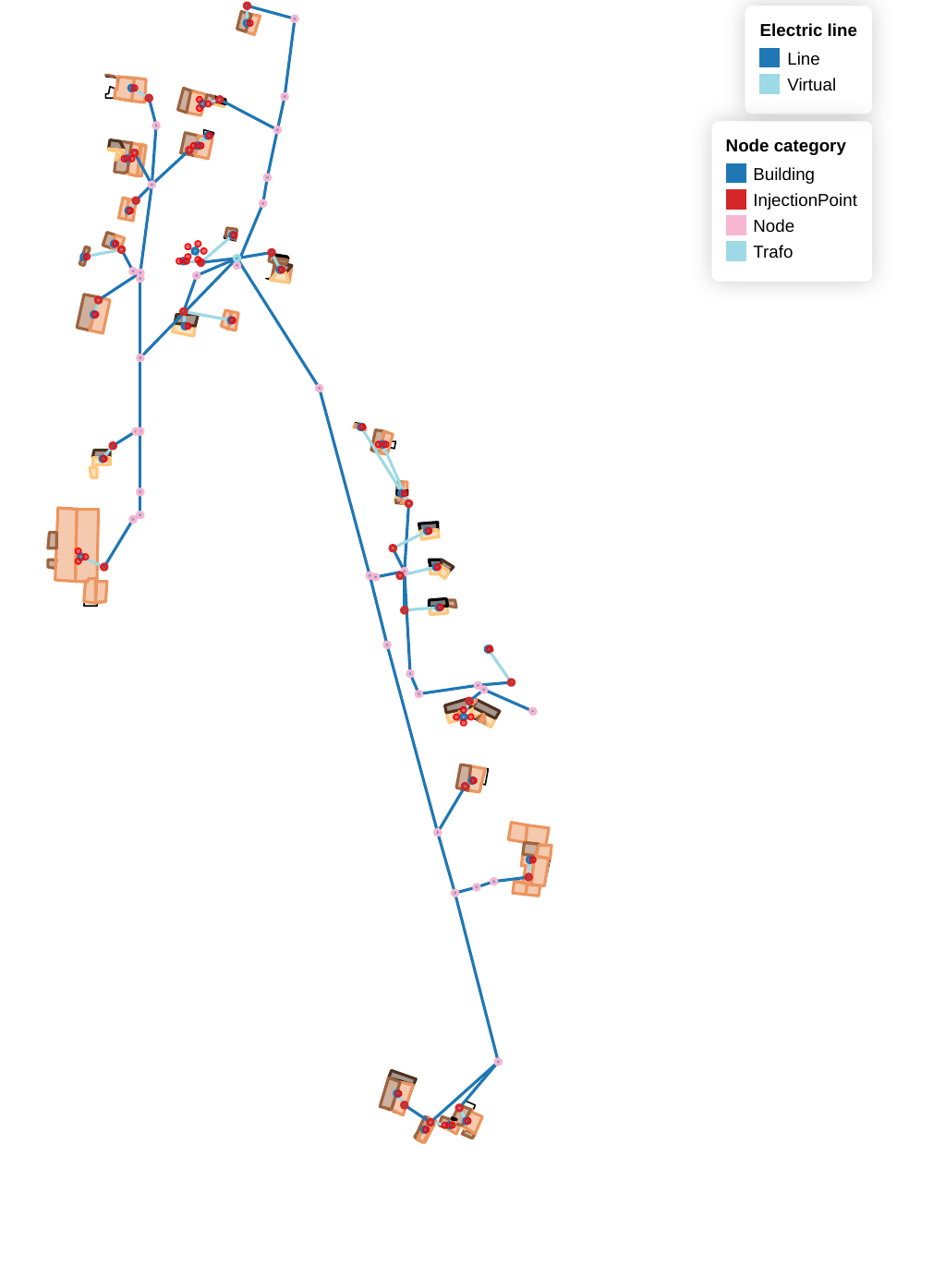}
    \caption{Rural LV network of the studied rural area.}
    \label{fig:grids}
\end{figure}

For the rural LV network, Romande Energie furnished details on the transformer, cables, and line characteristics, such as capacity, line length, and impedance. Additionally, the federal building identifier (EGID) of each customer linked to each injection point, as well as the measured load charge of the transformer station, was also provided. The LV network is displayed in Figure \ref{fig:grids}, and its main features are presented in Table \ref{tab:grids}.

\begin{table}[hbtp]
\centering
\caption{Main features of the rural LV network.}
\label{tab:grids}
\begin{tabular}{l c}
\hline
Feature & Value \\
\hline
Transformer rating & 630 kVA, 20 kV / 0.4 kV, 3-phase \\
Number of loads & 48 \\
Number of buildings & 32 \\
Number of injection points & 24 \\
Max power & 75.9 kW \\
Total consumption & 204.3 MWh/year \\
Cable type & Single-core copper, 4 AWG, PPP/PPL insulation \\
Line length and impedance & See~\ref{annex:line_parameters} for details \\
\hline
\end{tabular}
\end{table}

\subsection{Working assumptions}
Self-consumption behind the meter is assumed throughout the analysis, along with scenarios that either include or exclude shared storage, and involve adding or removing real-world actors. Additional modeling assumptions are specified as follows:

First, the optimization is performed at the individual building level. This allows the model to determine the optimal installation of PV and/or battery storage (if desirable) for each building. The optimization accounts for both Battery-to-Grid (B2G) and Grid-to-Battery (G2B) flows, intending to minimize total expenditure (TOTEX, see Eq. \ref{eq:obj}). The objective can also be set to maximize PV installation subject to the rooftop surface area available for each building. In this study, this configuration is adopted across all scenarios, including the baseline cases, which represent scenarios without a CEL, where each building operates independently and optimizes its own operation without any energy exchange between participants.

Second, the economic optimization of the CEL is carried out by aggregating all participating buildings into a single ``virtual building'' that represents the community as a whole. This virtual building is a purely conceptual entity used only for the economic analysis; it does not have a physical location in the grid. It combines the total electricity demand and PV generation of all CEL members by summing their respective 15-minute profiles, enabling the model to simulate the collective operation of the CEL (imports, exports, and shared storage) as a coordinated system. 

In contrast, the power flow analysis is conducted strictly at the individual building level (node level), preserving the full physical topology of the network. Although the CEL functions as a unified economic entity, it remains physically a collection of separate grid connection points. Therefore, the input for the power flow simulation includes the post-optimization load curves for each building. These curves have been updated to reflect the operational decisions made by the economic model, such as the dispatch of the centralized battery, effectively linking the community's flexibility strategies back to the specific physical nodes where the assets are located.

\subsection{Key performance indicators}
To evaluate the performance of CELs under the different scenarios, a set of key performance indicators (KPIs) is defined. These indicators capture both the economic outcomes, in terms of costs, savings, and profitability, and the technical outcomes, related to the operation of the low-voltage grid. The economic KPIs are quantified with explicit formulas, while the technical KPIs are derived from the results of the power flow analysis.

From the economic perspective, the indicators include the TOTEX (Eq.~\ref{eq:obj}), which combines OPEX and CAPEX over the 25-year lifetime of the system, as defined in Section~\ref{subsection:optimization_model}, the levelized cost of electricity (Eq.~\ref{eq:lcoe_system}), the annual total cost (Eq.~\ref{eq:total_cost}), the system profit expressed as the net present value of savings (Eq.~\ref{eq:profit}), the internal rate of return (Eq.~\ref{eq:irr}), and the revenue loss (Eq.~\ref{eq:revenue_loss}), which quantifies the decrease in DSO income caused by the formation of a CEL.

\begin{align}
\label{eq:lcoe_system}
\text{System LCOE}_b &= 
\frac{\sum_{t=1}^{T} \dfrac{C_{b,t}}{(1+r)^t}}{\sum_{t=1}^{T} \dfrac{L_{b,t}}{(1+r)^t}}
\end{align}
where $C_{b,t}$ is the investment plus operation and maintenance cost of building $b$ in year $t$, $L_{b,t}$ is the electricity supplied to its load in year $t$ (MWh), $r$ is the discount rate, and $T$ is the project lifetime (years). It gives a single value (in CHF/kWh) that reflects the overall economic efficiency of each building’s system.

\begin{align}
\label{eq:total_cost}
\text{Total Cost}_b &= C_{b,\text{power}} + C_{b,\text{energy}}
\end{align}
where $C_{b,\text{power}}$ covers fixed and capacity-related charges and $C_{b,\text{energy}}$ covers volumetric energy charges (CHF\,/\,year).

\begin{align}
\label{eq:profit}
\text{Profit}_b &= \sum_{t=0}^{T} \frac{C_{b,t,\text{baseline}} - C_{b,t,\text{scenario}}}{(1+r)^t}
\end{align}
where $C_{b,t,\text{baseline}}$ is the annual net cash flow of building $b$ when meeting the full load from the grid without PV or batteries, and $C_{b,t,\text{scenario}}$ is the corresponding net cash flow when PV and/or batteries are deployed. Thus, $\text{Profit}_b$ represents the Net Present Value (NPV) of the savings achieved by investing in local generation and storage, relative to a grid-only baseline. Each net cash flow accounts for investments, maintenance costs, and operating costs (grid imports minus grid exports). The discounted cumulative difference yields the reported profit.

\begin{align}
\label{eq:irr}
0 &= \sum_{t=0}^{T} \frac{R_{b,t} - C_{b,t}}{(1+\mathrm{IRR}_b)^t}
\end{align}
where $R_{b,t}$ is the net cash inflow (revenue) and $C_{b,t}$ the cash outflow (cost) for building $b$ in year $t$; the internal rate of return ($\mathrm{IRR}_b$) is the discount rate that sets the net present value to zero.

\begin{align}
\label{eq:bill_no_cel}
\text{Electricity bill without CEL}_b &= 
E_{b,\mathrm{energy}} + E_{b,\mathrm{tax}} + E_{b,\mathrm{grid}} \\
\label{eq:bill_with_cel}
\text{Electricity bill with CEL}_b &= 
E_{b,\mathrm{energy}} + E_{b,\mathrm{tax}} + E_{b,\mathrm{grid}} 
+ E_{b,\mathrm{energy}}^{\mathrm{CEL}} + E_{b,\mathrm{tax}}^{\mathrm{CEL}} + E_{b,\mathrm{grid}}^{\mathrm{CEL}} \\
\label{eq:revenue_loss}
\text{Revenue Loss}_b &= 
\text{Bill}_{b,\mathrm{no\,CEL}} - \left( \text{Bill}_{b,\mathrm{CEL}} - E_{b,\mathrm{energy}}^{\mathrm{CEL}} \right)
\end{align}

The electricity bill without a CEL (Eq.~\ref{eq:bill_no_cel}) represents the total annual payment for electricity imported from the main grid by building $b$, expressed in CHF/year. It includes all cost components: the energy price (\(E_{b,\mathrm{energy}}\)), taxes (\(E_{b,\mathrm{tax}}\)), and distribution charges (\(E_{b,\mathrm{grid}}\)).   

When a CEL is formed (Eq.~\ref{eq:bill_with_cel}), each building can also exchange electricity locally. Therefore, the total bill, also expressed in CHF/year, includes both the imported energy components (\(E_{b,\mathrm{energy}}\), \(E_{b,\mathrm{tax}}\), \(E_{b,\mathrm{grid}}\)) 
and the locally exchanged energy components (\(E_{b,\mathrm{energy}}^{\mathrm{CEL}}\), \(E_{b,\mathrm{tax}}^{\mathrm{CEL}}\), \(E_{b,\mathrm{grid}}^{\mathrm{CEL}}\)).  

Finally, the Revenue Loss (Eq.~\ref{eq:revenue_loss}) quantifies the decrease in total payments received by the DSO due to local energy exchanges within the CEL. It is computed as the difference between the total electricity bill in the baseline case (without CEL) and the bill when the CEL is active, excluding the locally exchanged energy component (\(E_{b,\mathrm{energy}}^{\mathrm{CEL}}\)). This adjustment ensures that only the part of the revenue that the DSO no longer collects - because the corresponding energy is exchanged locally - is considered a loss.

From the technical perspective, after performing the power flow analysis, we evaluate grid performance using the following indicators:

\begin{itemize}
    \item \textit{Maximum feed-in power}: the highest instantaneous active power injected into the grid by the CEL over the course of the year, indicating the potential stress on upstream transformers and lines due to PV surplus.
    \item \textit{Maximum power drawn}: the highest instantaneous active power imported from the grid to cover local demand in a year, which highlights peak demand conditions and their impact on transformer loading.
    \item \textit{Line loading percentages}: the ratio of the actual current flow in a line to its rated capacity, expressed in percent, used to identify potential line congestion or thermal overloading.
    \item \textit{Bus voltage deviations}: the deviation of nodal voltages from the nominal value (1 p.u.), which indicates whether over-voltage or under-voltage conditions occur within the network.
    \item \textit{Transformer voltage deviations}: the deviation of the transformer secondary voltage from its nominal value, used to assess whether voltage regulation at the point of common coupling remains within acceptable limits.
\end{itemize}

\subsection{Design of the Analysis}

To evaluate the economic and technical performance of CELs, we systematically varied a number of key variables that influence community operation, investment requirements, and impacts on the distribution grid. 

The first variable concerns the \textbf{community size and spatial distribution of its members}. Three configurations were analysed: \textit{CEL30} (9 buildings), \textit{CEL60} (19 buildings), and \textit{CEL100} (32 buildings), where the labels indicate the percentage of the 32 buildings that participate in the community. Beyond size, the spatial distribution of members within the feeder also plays a decisive role. To capture locational effects, two allocation strategies were considered. In the \textit{end-of-the-line} allocation, participants are deliberately placed at the electrically farthest end of the low-voltage network, a challenging case for grid operation due to longer electrical paths and more pronounced voltage deviations (both drops and rises) at the network boundaries. In the \textit{random allocation}, members are selected arbitrarily among the 32 available buildings, reflecting a realistic approach where community formation may occur without particular regard to electrical location. Because the random case yields a vast number of possible combinations (32!), capturing the full variability would require extensive sampling. As the most significant grid issues are expected at the end of the line, the main text therefore focuses on the end-of-the-line allocation, and the corresponding technical impacts are discussed in Section~\ref{sec:technical_analysis}. The selected results for the random allocation are reported in~\ref{annex:eco_results}.

The second variable is the level of \textbf{PV penetration} within the community, defined as the share of buildings equipped with a rooftop PV installation. Each system is sized to its maximum feasible capacity, corresponding to 70\% of the building’s available rooftop area. For instance, in CEL60, which includes 19 buildings, a 50\% penetration corresponds to 9 buildings hosting PV systems, while 100\% penetration indicates that all buildings are equipped. 

To test the robustness of the results, sensitivity analyses have been carried out for PV penetration levels of 25\%, 35\%, 50\%, 60\%, and 100\%. At very low penetration, too little surplus is expected for meaningful energy sharing, while at full deployment (100\%) internal exchanges are likely to decrease, limiting the added value of the CEL. The intermediate range of 25--60\% is therefore expected to offer the most relevant balance between self-consumption, intra-community trading, and grid interaction. Results for the 100\% PV case are included in the~\ref{annex:tech_results} and~\ref{annex:eco_results} for completeness.

A third variable concerns the \textbf{presence and location of community-scale batteries}. In this study, only a centralized battery for collective use is considered, while decentralized household storage is excluded. The sizing of this centralized battery is derived from the individual optimizations of each building: First, an optimal capacity is calculated for every building to minimize its TOTEX, and then the community-scale battery size is obtained by summing these individual capacities. The optimized capacities for the different CEL cases and PV penetration levels are summarized in Table~\ref{tab:battery_capacity}. The analysis, therefore, compares the distribution of energy with and without this collective battery, focusing on both its economic impact and its technical influence on the distribution grid.

\begin{table}[H]
\centering
\caption{Optimized community-scale battery capacities for each CEL case and PV penetration level.}
\resizebox{1\linewidth}{!}{%
\begin{tabular}{l
S[table-format=3.0]
S[round-mode=places,round-precision=0]
S[round-mode=places,round-precision=0]}
\hline
\textbf{Case} & \textbf{PV penetration (\%)} & \textbf{PV capacity (kW\textsubscript{P})} & \textbf{Battery Capacity [kWh]} \\
\hline
\multirow[c]{2}{*}{CEL100} 
 & 25 & 267.18 & 70.81 \\
 & 50 & 683.39 & 105.07 \\
\hline
\multirow[c]{2}{*}{CEL60} 
 & 25 & 98.30 & 32.71 \\
 & 50 & 285.77 & 56.12 \\
\hline
\multirow[c]{2}{*}{CEL30} 
 & 25 & 46.63 & 12.99 \\
 & 50 & 154.38 & 30.11 \\
\hline
\end{tabular}%
}
\label{tab:battery_capacity}
\end{table}

Two possible installation locations for this community-scale battery are investigated:  
\begin{itemize}
    \item \textit{Bat up}: installation at the site of the largest PV producer, close to the transformer.  
    \item \textit{Bat down}: installation at the end of the line, closer to end users but still close to the major PV producer.  
\end{itemize}

For the assessment of grid impacts, the \textit{Bat down} configuration is adopted in all cases, except for the baselines without batteries and the dedicated \textit{Bat up} scenario. To provide a sense of scale, the maximum community storage capacity considered in this study (approx. 105~kWh) is roughly equivalent to the battery capacity of two standard electric vehicles or a single high-performance one.

Finally, the analysis also considers cases with \textbf{large consumers or producers} to capture situations where the community composition is strongly heterogeneous. These scenarios test the resilience of the CEL framework under asymmetrical conditions, where a single actor can substantially alter the balance between demand and supply.  

To simulate such situations, a \textit{large consumer} is introduced with an annual demand of 22~MWh, roughly four times higher than the average building consumption of 6.3~MWh. This value corresponds to the annual demand for the building that hosts the largest PV producer in the network. In parallel, a \textit{large producer} case is modeled by adding a PV generation profile from a 172.3~kW\textsubscript{P} installation, which represents the largest PV producer in the network. Since these scenarios are part of the economic analysis, the additional load and generation profiles are incorporated into the aggregated CEL representation (the virtual building), without being assigned to a specific physical location in the grid. Detailed building-level data and PV capacities are provided in~\ref{annex:building_data}.

Together, the combination and variation of these variables provide a comprehensive framework for evaluating the performance of CELs under different conditions, ensuring that both representative configurations and edge cases are addressed in the analysis.

\section{Results}
\subsection{Technical Analysis}
\label{sec:technical_analysis}
This section presents the technical evaluation of power and energy flows, congestion risks, and network stability across the simulated scenarios. The analysis is structured into three parts: transformer loading, voltage deviations, and line loading. All scenarios, except for the baselines, include a centralized battery sized optimally for each CEL. The focus is placed on CELs located at the end of the line, specifically CEL30 and CEL60, while results for CEL100 are provided in~\ref{annex:eco_results}. It is worth noting that the analysis covers the entire low-voltage network and not only the buildings within the CEL.

\paragraph{Transformer Loading Analysis}
Table~\ref{tab:trafo_load} presents the yearly maximum transformer-level feed-in and drawn power, expressed in kW and as percentages relative to the corresponding baseline values for each CEL size. These maximum power values are identical for both the double tariff and the dynamic tariff scenarios because the external tariff used is the same regardless of the CEL's internal pricing scheme, thus leading to identical transformer loading results, which is why the results are only presented once in the table.

\begin{table}[h!]
\centering
\caption{Yearly maximum feed-in and drawn power values by the buildings forming the CEL, evaluated at the transformer level for different CEL scenarios, expressed in kW and as percentages relative to the baseline values of each CEL size. The baseline corresponds to the case without a CEL and with 100\% PV penetration.}
\label{tab:trafo_load}
\renewcommand{\arraystretch}{1.2}
\begin{adjustbox}{max width=\textwidth}
\begin{tabular}{l *{4}{S[table-format=4.1]}}
\toprule
Name
& {Feed-in power [kW]} & {Reference \%}
& {Power drawn [kW]} & {Reference \%} \\
\midrule
Baseline CEL60   & 462.1 & 100.0 & 81.5  & 100.0 \\
CEL60 max PV     & 462.1 & 100.0 & 119.6 & 146.8 \\
CEL60 50PV       & 213.6 & 46.2  & 118.0 & 144.8 \\
CEL60 25PV       & 58.0  & 12.6  & 91.7  & 112.5 \\
\midrule
Baseline CEL30   & 263.5 & 100.0 & 81.5  & 100.0 \\
CEL30 max PV     & 263.5 & 100.0 & 92.2  & 113.2 \\
CEL30 50PV       & 97.7  & 37.1  & 88.5  & 108.6 \\
CEL30 25PV       & 20.7  & 7.9   & 81.5  & 100.0 \\
\bottomrule
\end{tabular}
\end{adjustbox}
\end{table}

Regarding feed-in power, results follow the expected trend: higher PV penetration leads to increased feed-in peaks. Notably, for the maximum PV cases, these maximum values remain unchanged from the baseline, despite the addition of a battery, indicating that the economically optimal storage capacity is relatively small compared to the installed PV capacity, and therefore insufficient to absorb the significant generation surplus during peak PV generation. However, regarding the power drawn from the grid, a key finding is that the maximum peak paradoxically increases in scenarios with a centralized battery under both tariff structures. For CEL60, high PV penetration scenarios (50\% and 100\%) reach 118--120~kW (i.e., $145$--$147\%$ of the baseline), while CEL30 peaks reach 92~kW (113\%). This new, higher peak is not just driven by building demand, but by the battery's economic optimization strategy.

\begin{figure}[H]
\centering
\includegraphics[width=0.8\textwidth]{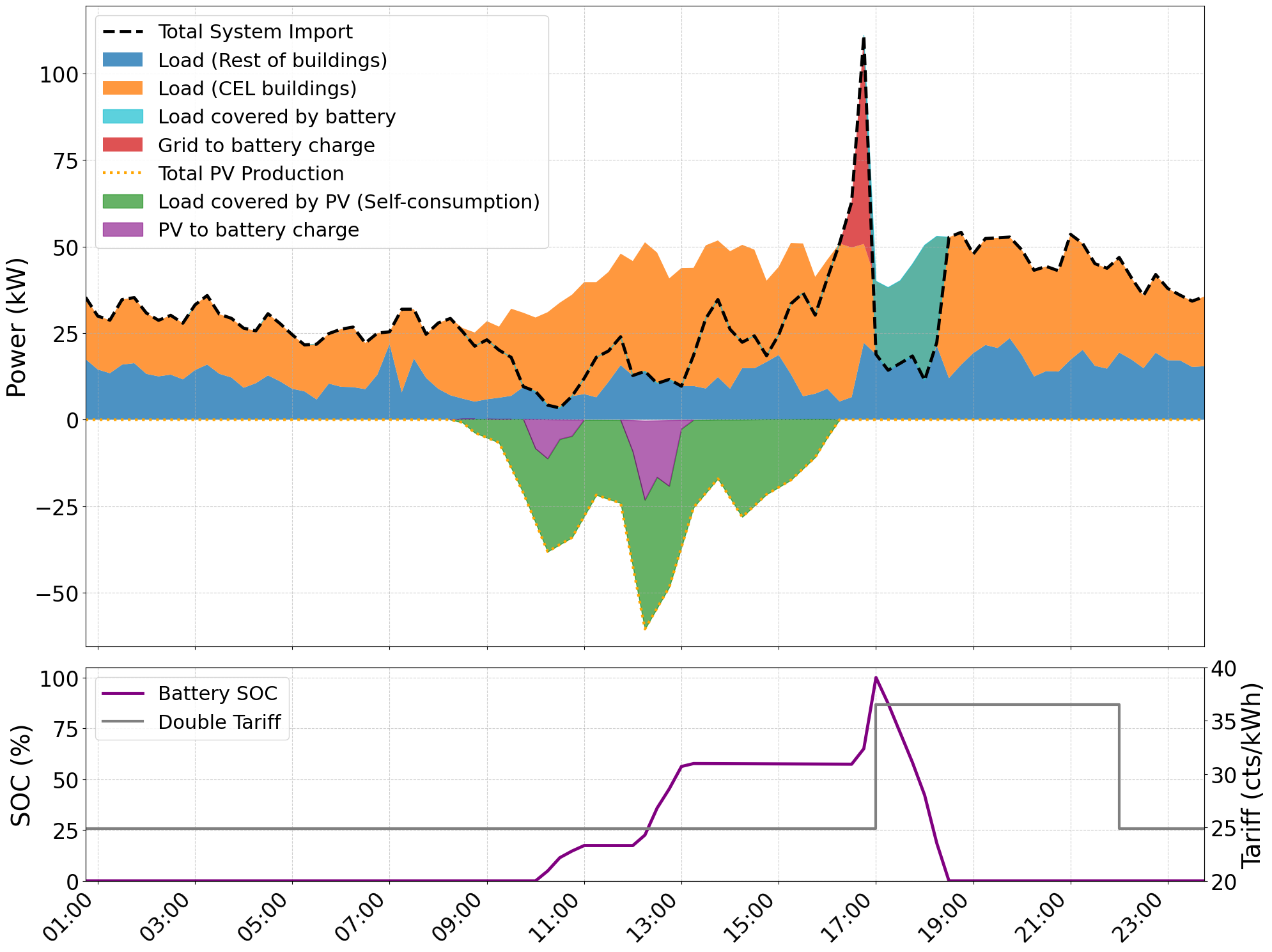}
\caption{System power flows (top) and battery State of Charge alongside the double tariff profile (bottom) on a representative day (25 November).}
\label{fig:peak_day_analysis}
\end{figure}

Figure~\ref{fig:peak_day_analysis} provides a detailed illustration of this phenomenon, showing the system's power flows on a representative day where this peak occurs. The graph clearly shows the existing building loads (blue and orange areas). During the day, the battery is partially charged using available solar energy (as seen in the PV to battery charge area, colored purple). However, on this late-autumn day, the solar charging is insufficient to fill the battery, resulting in a low State of Charge (SOC) by late afternoon (as shown in the bottom panel). Critically, once the PV production ceases (after 16:00) and immediately before the 17:00 external tariff change (indicated by the step in the grey line on the bottom plot), a massive battery charge from the grid (red area) begins. This is a deliberate action by the optimizer to fill the remaining capacity with less expensive off-peak power before the more expensive peak period begins. This battery charging demand then stacks directly on top of the existing building load, creating the new, artificial system peak (the black dashed line). Although this behavior can be seen as a model artifact, it also reflects the possible problems that can be encountered with limited forecasting and boundary-driven charging, due to changes in the tariff.

As shown in Table~\ref{tab:trafo_load}, the results for the dynamic tariff scenarios are identical to those of the double tariff. This is because the tariff structure applied to energy imported from the external grid remains the double tariff in both configurations. The distinction between the scenarios lies only in the internal tariff used for billing flows between building members. Since the internal tariff effect is purely economic and does not modify the external cost signal or physical constraints, the battery executes the same arbitrage strategy in both cases. Consequently, the physical power flows remain unchanged, resulting in the same peak loads for both tariff schemes.

\begin{figure}[h]
\centering
\includegraphics[width=0.8\textwidth]{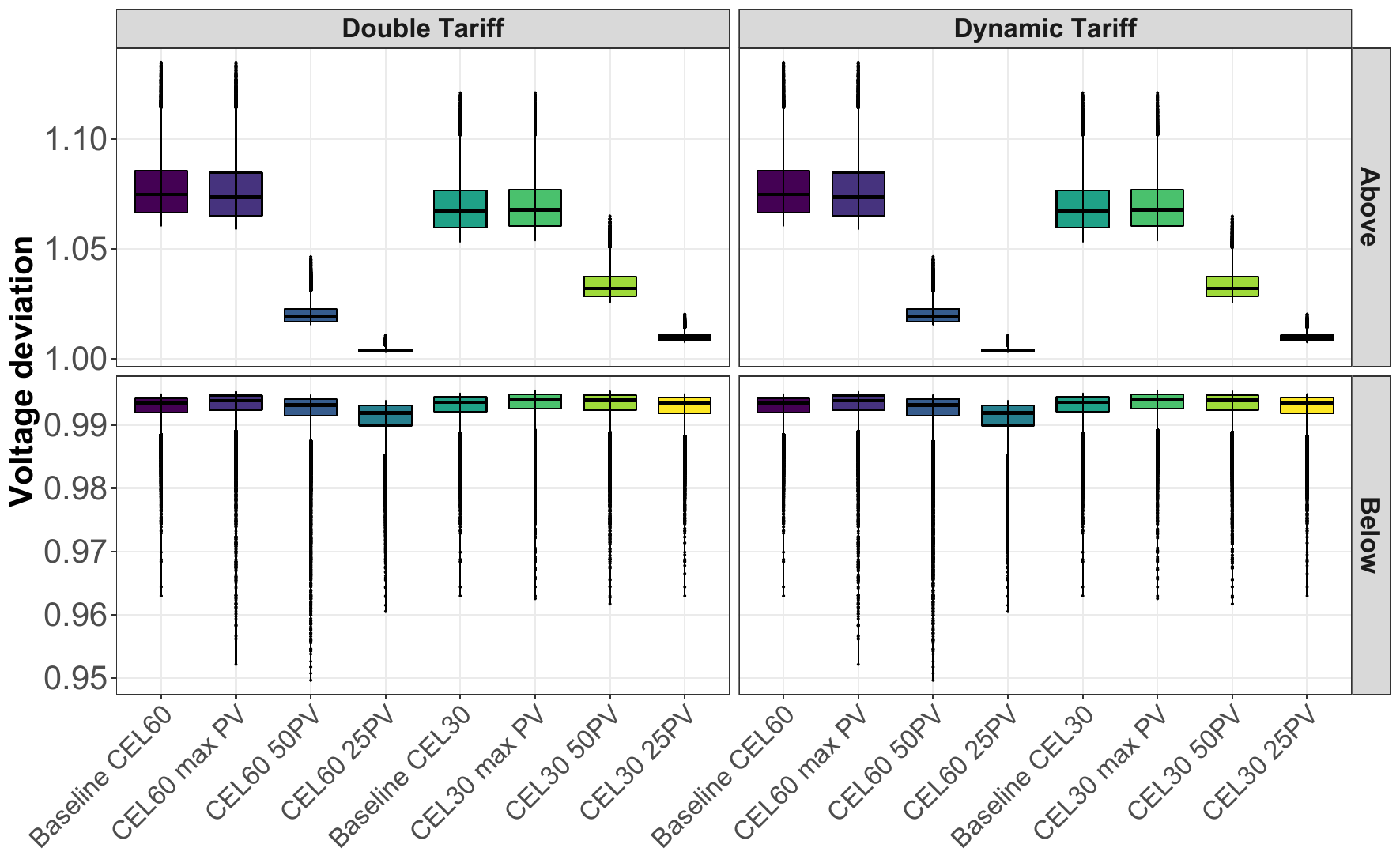}
\caption{Voltage deviation distribution (95th percentile) for CEL30 and CEL60 under double and dynamic tariffs. The baseline scenarios correspond to the cases without a CEL, with 100\% PV capacity and no battery storage. The max-PV, 50\% PV, and 25\% PV cases include an economically optimal centralized battery sized for the CEL.}
\label{fig:voltage_cel30_60}
\end{figure}

\paragraph{Voltage Deviation Analysis}
Figure~\ref{fig:voltage_cel30_60} presents the voltage deviation distributions for CEL60 and CEL30 scenarios using box plots of the 95th percentile values, under varying levels of PV penetration and the two tariffs. The 95th percentile corresponds to the voltage level that is exceeded in only 5\% of the time steps, thus highlighting severe deviations without being dominated by a single extreme value. Each box displays the inter-quartile range (IQR) with the median marked by a horizontal line, whiskers extending up to 1.5 times the IQR, and points representing outliers. The top panel displays over-voltages ($V > 1.0$ p.u.; nominal = 1.0 p.u.), and the bottom shows under-voltages ($V < 1.0$ p.u.).

The baseline scenario corresponds to all buildings within the CEL having 100\% PV capacity installed but no battery storage. In contrast, the max PV cases (e.g., CEL60 max PV, CEL30 max PV) include an economically optimal centralized battery sized for the CEL. As shown in Figure \ref{fig:voltage_cel30_60}, the reduction of over-voltages is negligible, with the median voltage differing by only a few hundredths of a unit. Specifically, the CEL60 baseline median in the double-tariff panel is about 1.07 p.u., and the CEL60 max-PV case is roughly 0.01 p.u. lower. For CEL30, the baseline median is about 1.06 p.u., and in the max-PV case, it is only marginally lower, with the difference hardly visible in the plot.

Over-voltage is the primary challenge, occurring when PV generation exceeds local consumption, causing reverse power flow and voltage rise. At high PV penetration (CEL60 max PV and CEL30 max PV), voltages reach 1.10-–1.12 p.u, exceeding operational limits. Reducing penetration to 50\% and 25\% progressively mitigates this. The dynamic tariff yields over-voltages essentially identical to the double tariff.

Under-voltage is far less severe: medians remain close to nominal (approximately 0.99 p.u.) with few outliers, indicating robust performance against voltage drops. As expected and indicated in previous studies \cite{penabello2024arxiv}, the main issue is PV-induced voltage rise, not load-induced drops.

CEL60 generally achieves better over-voltage mitigation than CEL30, except in the max-PV case. In the max-PV scenarios, CEL60 shows slightly higher voltages (median  $\approx 1.07$ p.u., upper values up to 1.11–1.12 p.u.) compared to CEL30 (median  $\approx 1.06$ p.u., upper values around 1.10–1.11 p.u.). By contrast, at 50\% PV penetration, CEL60 maintains voltages closer to 1.02 p.u., whereas CEL30 is nearer to 1.03 p.u.; at 25\% PV, both CELs converge to about 1.01 p.u., with CEL60 being slightly lower. The higher over-voltage in CEL60 at full PV penetration results from its larger total installed capacity, which increases simultaneous injections into the network. However, at 25 - 50\% PV, CEL60's larger and more diverse member base, together with its higher battery capacity, helps absorb generation peaks and smooth net flows, leading to lower voltages than in CEL30.

\paragraph{Line Loading Analysis}
The analysis of line loading confirms that the establishment of a CEL does not inherently change the physical power flows within the network, as all exchanges occur through the public distribution grid. Instead, variations in line utilization are driven by the location and operation of shared assets, particularly the PV installations and the central battery. Comparing the baseline with the CEL100 battery-up and battery-down cases shows that the impact of storage on line loading depends strongly on its placement within the network.

For instance, as summarized in Table~\ref{tab:line_loading_battery_position}, some lines experience slightly lower loadings when the battery is placed closer to the transformer (e.g., line~46), while others show higher utilization when the battery is relocated at the end of the line (e.g., line~40). In extreme cases, the median loading of line~40 increases from 41.4\% in the baseline to 43.3\% when the battery is positioned closer to the transformer, indicating that non-optimal siting can lead to marginal increases in loading on specific branches. Similar patterns are observed for CEL60, where line~39 shows an increase from 28.5\% to 29.6\% under the max-PV case.

To provide a complete overview, Tables~\ref{tab:most_loaded_max} and~\ref{tab:most_loaded_median} in the~\ref{annex:tech_results} report the maximum and median loading of the six most stressed lines across all scenarios. Since these tables are extensive, Table~\ref{tab:line_loading_battery_position} presents a subset of representative results to illustrate how battery location influences network utilization.

\begin{table}[H]
\centering
\caption{Effect of battery location on median line loading (\%) for selected scenarios under the double tariff.}
\label{tab:line_loading_battery_position}
\renewcommand{\arraystretch}{1.1}
\setlength{\tabcolsep}{6pt}
\begin{tabular}{lccc}
\toprule
\textbf{Line} & \textbf{Baseline} & \textbf{CEL100 bat up} & \textbf{CEL100 bat down} \\
\midrule
40 & 41.4 & 43.3 & 41.6 \\
46 & 39.4 & 35.6 & 39.6 \\
39 & 35.0 & 35.1 & 36.4 \\
\bottomrule
\end{tabular}
\end{table}

Tariff comparison confirms that double and dynamic tariffs produce similar maximum line loadings, suggesting that tariffs with only a volumetric structure have little influence on extreme peaks. However, there is a minor exception on critical lines where the internal dynamic price alters battery behavior. For example, for line 39 in the CEL100 50PV scenario, the maximum loading decreases from 38.4\% (double tariff) to 34.5\% (dynamic tariff), and for the same line in CEL100 25PV, it falls from 29.3\% to 27.4\%. Although these reductions are localized and the absolute differences remain modest, they suggest that dynamic pricing offers a slight advantage in managing specific peak loading events compared to the double tariff.

\subsection{Economic Analysis}
We first examine the economic assessment of the case in which every building installs rooftop PV at its maximum feasible capacity. Although this scenario maximizes the potential for renewable generation, it quickly proved to be economically unattractive, resulting in the lowest IRR among all CEL scenarios (9.3\%). With all members producing electricity simultaneously, opportunities for mutually beneficial exchanges within the CEL are minimal, and most of the PV surplus must be exported to the grid at low feed-in tariff, also creating challenges for the network.

\subsubsection{Baseline and Community Configurations}
Table~\ref{tab:lcoe_cel3060_bat} compares the baseline configuration, where buildings operate individually without forming a CEL, to community cases with 25\% and 50\% PV penetration, with and without a centralized battery. In the baseline, the LCOE corresponds to the average of the individual LCOEs of all buildings and reflects stand-alone investments without (remunerated) energy exchanges within the LV network. The results show that joining a CEL consistently reduces electricity costs for its members and increases profitability compared to the baselines, highlighting the financial advantage of collective energy management.

\begin{table}[htbp]
\centering
\caption{Techno-economic results for CEL30 and CEL60 at PV penetration levels of 25\% and 50\%, including the Levelized Cost of Electricity (LCOE), Internal Rate of Return (IRR), and total profit over a 25-year lifetime. The baselines correspond to each PV penetration level without a CEL, while additional scenarios include a CEL with or without a centralized battery.}
\resizebox{\linewidth}{!}{%
\begin{tabular}{ll
S[round-mode=places,round-precision=4] 
S[round-mode=places,round-precision=1] 
S[round-mode=places,round-precision=1]}
\hline
\textbf{Case} & \textbf{Scenario} & \textbf{LCOE (CHF/kWh)} & \textbf{IRR (\%)} & \textbf{Profit (kCHF)} \\
\hline
\multirow{4}{*}{CEL60} 
& Baseline 50\% PV   & 0.220638 & 7.380194847 & 283.178 \\
 & 50\% PV     & 0.1017 & 10.9278677857822 & 375.79 \\
 & 50\% PV + Bat    & 0.0943 & 11.047694441302 & 392.602 \\
 & Baseline 25\% PV   & 0.189398741 & 13.06489387 & 94.981 \\
 & 25\% PV     & 0.1868 & 13.2326268451571 & 182.56 \\
 & 25\% PV + Bat    & 0.1826 & 13.313824924638 & 192.163 \\
\hline
\multirow{4}{*}{CEL30} 
 & Baseline 50\% PV     & 0.122105642 & 12.29786101 & 161.916 \\
 & 50\% PV     & 0.1198 & 12.5935808324359 & 239.763 \\
 & 50\% PV + Bat    & 0.1136 & 12.7022014554216 & 249.851 \\
 & Baseline 25\% PV     & 0.233862068 & 13.83053725 & 49.442 \\
 & 25\% PV     & 0.2038 & 14.8260115760244 & 103.001 \\
 & 25\% PV + Bat    & 0.2012 & 14.866726493271 & 107.2 \\
\hline
\end{tabular}%
}
\label{tab:lcoe_cel3060_bat}
\end{table}

For CEL60, community participation significantly improves economic performance. At 50 \% PV penetration, the LCOE drops from 0.221 CHF/kWh in the baseline to 0.102 CHF/kWh in the CEL configuration, while lifetime profits increase from 283 kCHF to nearly 376 kCHF. The inclusion of a shared battery further reduces the LCOE to 0.09 CHF/kWh and raises profits to about 393 kCHF, confirming that collective investment and local energy use deliver tangible financial benefits.

For CEL30, a similar pattern is observed. At 50\% PV penetration, the LCOE decreases from 0.122 to 0.120~CHF/kWh, and profits rise from 162 kCHF to about 240 kCHF. With the addition of a battery, the LCOE falls slightly to 0.11 CHF/kWh and profits reach around 250 kCHF. Although the absolute gains are smaller than in CEL60 due to its smaller scale, both cases confirm that coordinated operation within a CEL substantially improves cost efficiency and investment returns compared to their respective baselines, with the inclusion of shared batteries further enhancing overall economic performance.

\subsubsection{Effect of PV Penetration and Battery Integration}
In general, the results show that intermediate PV penetration provides the most favorable balance between relative profitability and absolute profits. At 25\%, both CEL60 and CEL30 reach higher IRRs but only modest cumulative gains, while at 50\% penetration the IRRs remain attractive and the absolute profits grow substantially. The addition of a community-scale battery brings only marginal improvements, since the optimal storage size is relatively small, especially in CEL30. This confirms that under the current tariff design, profitability is primarily driven by PV penetration rather than storage.

\begin{figure}[H]
\centering
\includegraphics[width=0.75\textwidth]{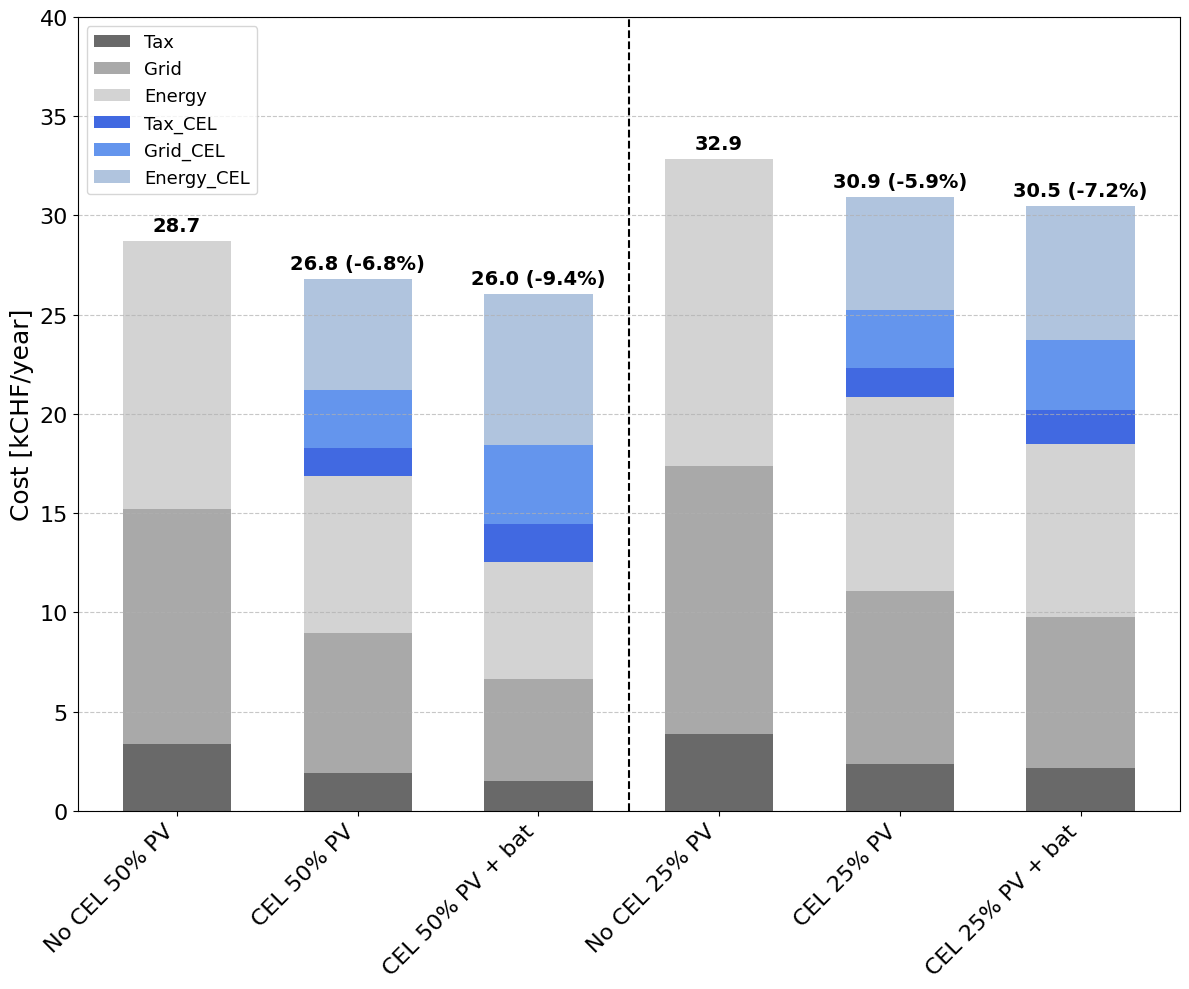}
\caption{Yearly electricity bill for CEL60 for two PV penetration levels (25\% and 50\%) and three different scenarios: PV only (no CEL), PV with a CEL, and a CEL with PV and a centralized battery.}
\label{fig:bill_cel60}
\end{figure}

A complementary perspective on economic outcomes is given by the annual electricity bills, shown in Figure~\ref{fig:bill_cel60} for CEL60. Each bar is divided into three main components: taxes, distribution charges, and energy costs. In the cases with a CEL, three additional elements are present, to account for the energy exchanged within the community 
(\textit{Tax\_CEL}, \textit{Grid\_CEL}, and \textit{Energy\_CEL}). This representation highlights how local exchanges affect the cost structure. 

The results show that forming a CEL consistently reduces the overall annual bill compared to the non-CEL baselines. The largest savings are derived from the distribution component, as internal exchanges incur lower network charges. In contrast, the tax component remains unchanged. The impact of PV penetration is particularly evident: at 25\% PV, total costs fall by around 6--7\%, while at 50\% PV the reductions become much more significant, reaching up to 9\% when storage is added.  This suggests that an intermediate penetration level around 50\%, where annual PV generation is approximately 2.22 times the community’s total demand, constitutes a favorable operating point, where the community benefits the most from both self-consumption and internal exchanges. Adding a battery further enhances the reduction in costs, although the difference compared to the PV-only case remains moderate (about 2–3\%), the reduction relative to the non-CEL configuration reaches up to 9.4\%.

For CEL30, the results follow the same trends, with reductions of similar magnitude across all cases. To avoid redundancy, these results are reported in the~\ref{annex:eco_results} (Figure~\ref{fig:bill_cel30}), which confirms the robustness of the conclusions in different community sizes.

\begin{figure}[htbp]
\centering
\includegraphics[width=0.8\textwidth]{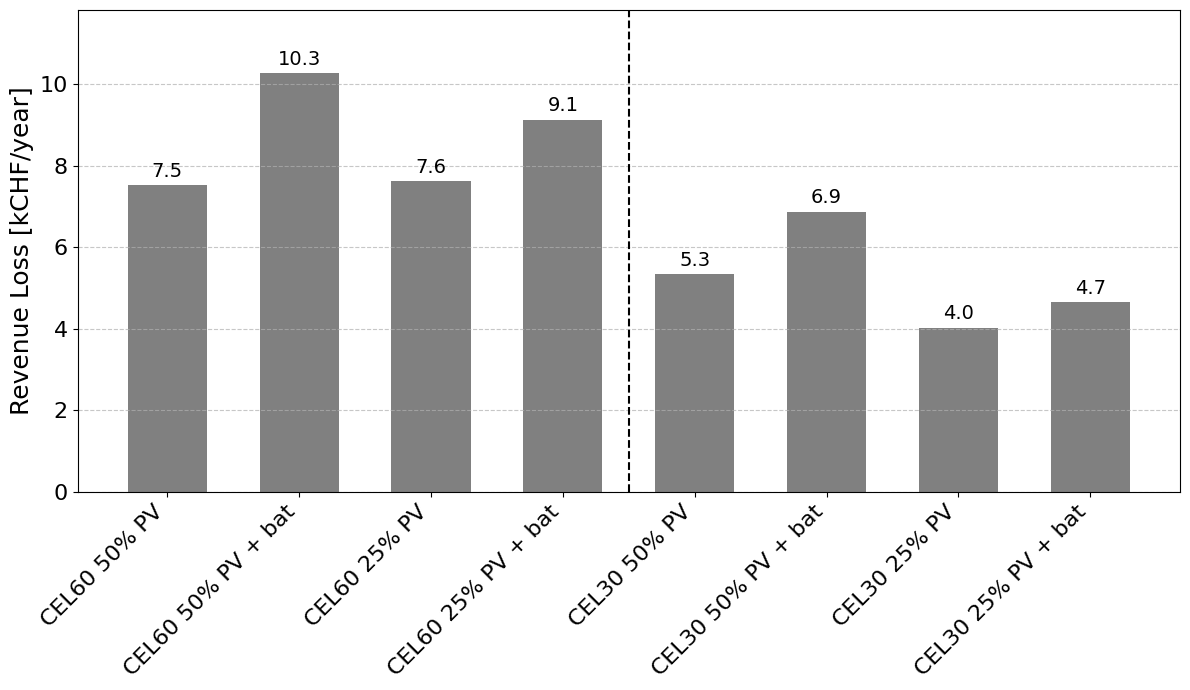}
\caption{Annual revenue loss for the DSO due to CEL formation, for different community sizes (CEL30 and CEL60) and PV penetration levels (25\% and 50\%), with and without a centralized battery.}
\label{fig:rev_loss}
\end{figure}

An additional way to interpret the economic effects of CEL formation is given in Figure~\ref{fig:rev_loss}, which shows the difference in revenue losses between CEL and non-CEL cases. This indicator captures the monetary impact of local exchanges from the DSO’s perspective. Revenue losses occur primarily because CEL members import less electricity from the grid, a direct consequence of self-consuming local PV production and internal energy exchange. As this energy is settled financially between members rather than with the DSO, the grid operator sees a reduction in payments for energy, distribution use, and taxes. It can thus be interpreted in two complementary ways: as a gain from the perspective of the community members, or as a revenue loss from the perspective of the DSO.

The results highlight that DSOs face a non-negligible reduction in revenues when CELs are formed. For CEL60, the losses range from 7.5~kCHF/year at 50\% PV (equivalent to 26.2\% of the total annual bill) to over 10~kCHF/year when a battery is added (35.8\%), while at 25\% PV the losses remain between 7.6 and 9.1~kCHF/year (23.2\%--27.8\%). In CEL30, the absolute values are smaller, between 4.0 and 6.9~kCHF/year (17.6\%--33.5\%), depending on PV penetration and storage. These values illustrate that although CELs reduce participant costs and enhance local consumption, they also imply systematic reduction in DSO revenues, which may require regulatory attention to ensure fair cost recovery for DSOs.

However, the larger revenue losses observed in CEL60 do not imply that a single large CEL performs better than smaller ones. This difference mainly reflects the aggregation of more buildings in CEL60, which results in higher overall demand and PV generation, and consequently greater internal energy exchanges that reduce grid imports and DSO revenues. In addition, the specific composition of CEL60, in terms of its mix of consumers and producers, amplifies internal trading. Therefore, the contrast between CEL30 and CEL60 should be interpreted as a scale and composition effect rather than as evidence that larger CELs are more advantageous.
\newpage
\subsubsection{Lifetime Cost Structure: TOTEX, CAPEX, and OPEX}
\begin{figure}[htbp]
\centering
\includegraphics[width=0.8\textwidth]{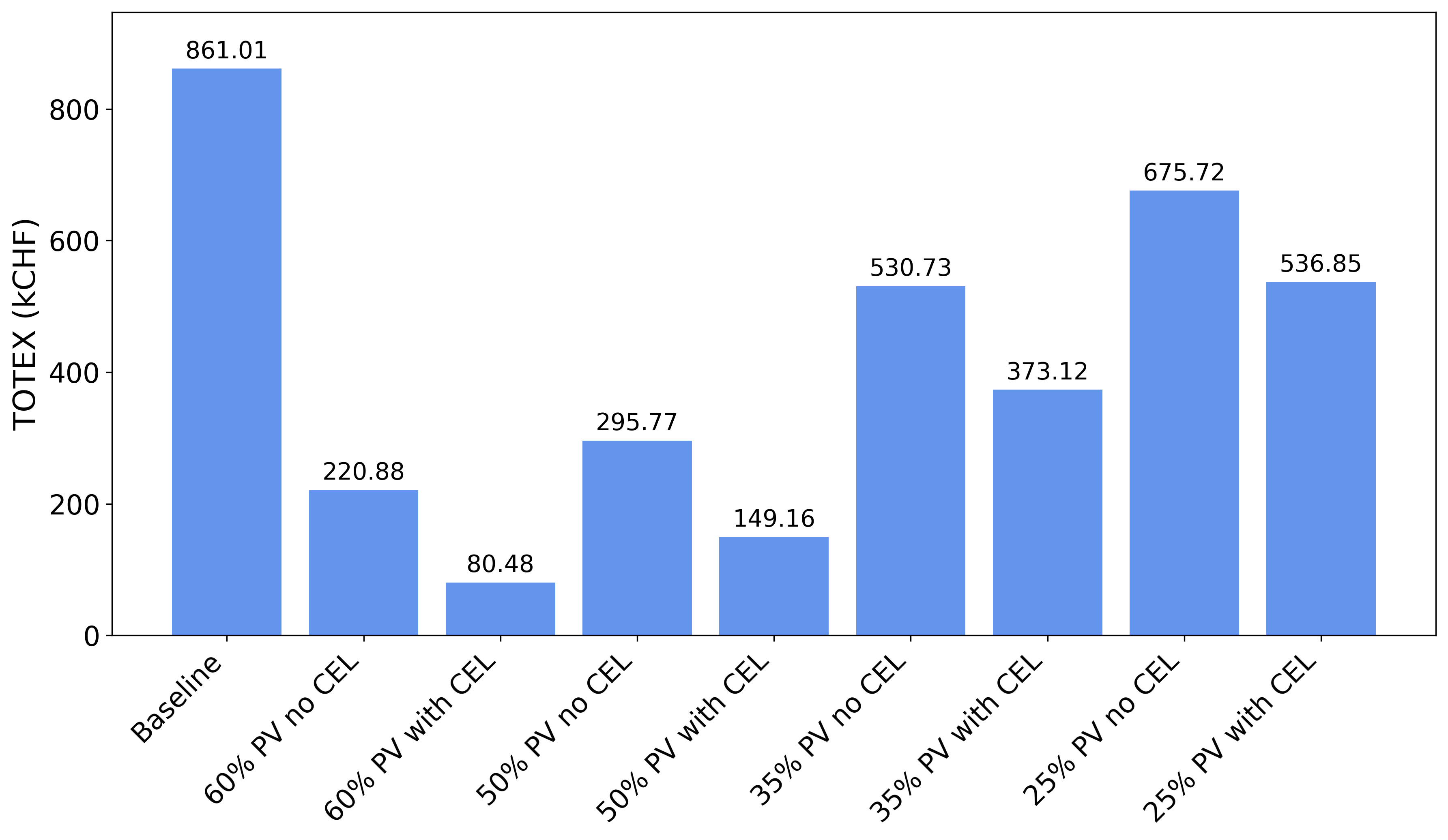}
\caption{TOTEX for CEL60 over a 25-year lifetime, comparing a baseline case without PV and without a CEL to CEL60 configurations with different levels of PV penetration.}
\label{fig:totex_cel60}
\end{figure}

Beyond yearly expenditures, it is also important to examine the cost structure for the entire life of the project. To this end, the analysis was extended to the TOTEX, CAPEX, and OPEX values. Figure~\ref{fig:totex_cel60} illustrates the results for CEL60, while the corresponding figure for CEL30 can be found in  Figure~\ref{fig:totex_cel30} of the~\ref{annex:eco_results}. All values are expressed for a 25-year horizon, with OPEX including both maintenance and equipment replacement. The baseline corresponds to the case in which all energy demand is supplied from the grid without any PV or community exchanges.

The results show that the formation of a CEL consistently reduces total lifetime costs compared to the non-CEL case at the same PV penetration level. As PV capacity increases, TOTEX continues to decrease, but the profitability of the investment, expressed through the IRR, declines progressively. In other words, more PV makes energy cheaper (lower LCOE) but reduces the return on investment due to the higher upfront costs and diminishing marginal savings. As it can be seen in Table~\ref{tab:lcoe_cel3060}, for CEL60, the LCOE falls from 0.1868 CHF/kWh at 25 \% PV to 0.0525 CHF/kWh at 80 \%, while the IRR drops from 13.23 \% to 9.69 \%. The same tendency is observed in CEL30, although the variability is more pronounced due to its smaller number of participants. The slight increase in LCOE at 60 \% PV for CEL30 (compared to 50 \%) arises from the inclusion of an additional building with a very low annual demand (206 Wh) and an oversized PV installation (2 kWp) compared to its consumption. This weak load-generation balance lowers the economic efficiency of the system.

\begin{table}[H]
\centering
\caption{Techno-economic results for CEL30 and CEL60 cases under different PV penetration levels, including the Levelized Cost of Electricity (LCOE), Internal Rate of Return (IRR), and total profit over a 25-year lifetime.}
\resizebox{\linewidth}{!}{%
\begin{tabular}{ll
S[round-mode=places,round-precision=4] 
S[round-mode=places,round-precision=2] 
S[round-mode=places,round-precision=1]}
\hline
\textbf{Case} & \textbf{PV penetration (\%)} & \textbf{LCOE (CHF/kWh)} & \textbf{IRR (\%)} & \textbf{Profit (kCHF)} \\
\hline
\multirow{7}{*}{CEL60} 
 & 25 & 0.1868 & 13.2326 & 182.560 \\
 & 35 & 0.1500 & 11.9854 & 266.288 \\
 & 50 & 0.1017 & 10.9279 & 375.790 \\
 & 60 & 0.0894 & 10.3718 & 403.825 \\
 & 65 & 0.0753 & 10.2091 & 435.820 \\
 & 70 & 0.08 & 9.8446 & 425.281 \\
 & 80 & 0.0603 & 9.5793 & 469.913 \\
\hline
\multirow{7}{*}{CEL30} 
 & 25 & 0.2038 & 14.8260 & 103.001 \\
 & 35 & 0.1424 & 13.5549 & 202.870 \\
 & 50 & 0.1198 & 12.9358 & 239.763 \\
 & 60 & 0.1245 & 11.8055 & 232.090 \\
 & 70 & 0.11369 & 11.2373 & 249.770 \\
 & 80 & 0.0748 & 10.9801 & 312.824 \\
\hline
\end{tabular}%
}
\label{tab:lcoe_cel3060}
\end{table}

\begin{figure}[H]
\centering
\includegraphics[width=0.7\textwidth, height=0.57\textheight]{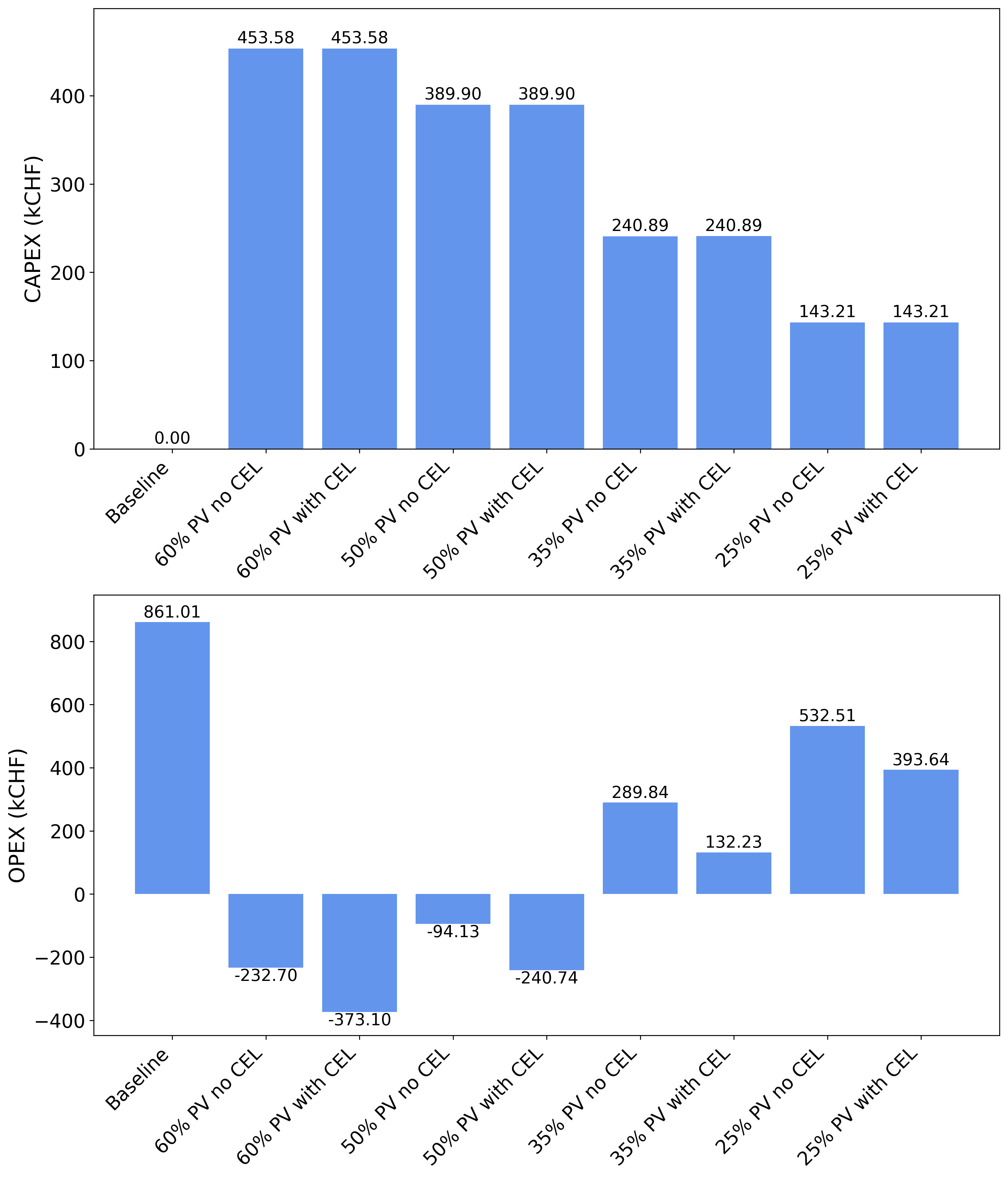}
\caption{Decomposition of TOTEX into CAPEX and OPEX for CEL60, comparing a baseline case without PV and without a CEL to CEL60 configurations with different PV penetration levels. OPEX includes maintenance and equipment replacement.}
\label{fig:capex_opex_cel60_partial}
\end{figure}

To better understand the cost structure, Figure~\ref{fig:capex_opex_cel60_partial} shows the decomposition into CAPEX and OPEX for CEL60. Higher PV penetration requires larger upfront investments, reflected in higher CAPEX. However, these investments are more than compensated by sharp reductions in OPEX, which even become negative in some CEL cases, meaning that operational revenues exceed maintenance and replacement costs.

Overall, the results confirm that PV penetration remains the main driver of both the cost structure and profitability. Increasing PV capacity gradually shifts the cost balance from OPEX to CAPEX, reducing energy costs but yielding lower financial returns. Finding the right balance between these two effects is thus essential for designing economically viable and scalable CELs.

\subsubsection{Effect of Community Composition}
In addition to the sensitivity analysis on PV penetration, the composition of the community was also found to play a decisive role. Beyond the standard CEL configurations, additional cases were analyzed to assess how the inclusion of heterogeneous members affects community performance. Two types of participants were considered: a \textit{large consumer} (22.1~MWh annual demand without PV) and a \textit{large producer} (172~kWp of installed PV without demand). Instead of selecting buildings with the smallest PV systems, which also had very low electricity consumption and therefore limited potential for meaningful exchanges, the choice of participants was refined to include those with the lowest PV-to-consumption ratios. This ensured that the added members could actively contribute to internal energy redistribution, maximizing the potential for local exchange within the CEL. 

\begin{figure}[H]
    \centering
    \includegraphics[width=1\textwidth]{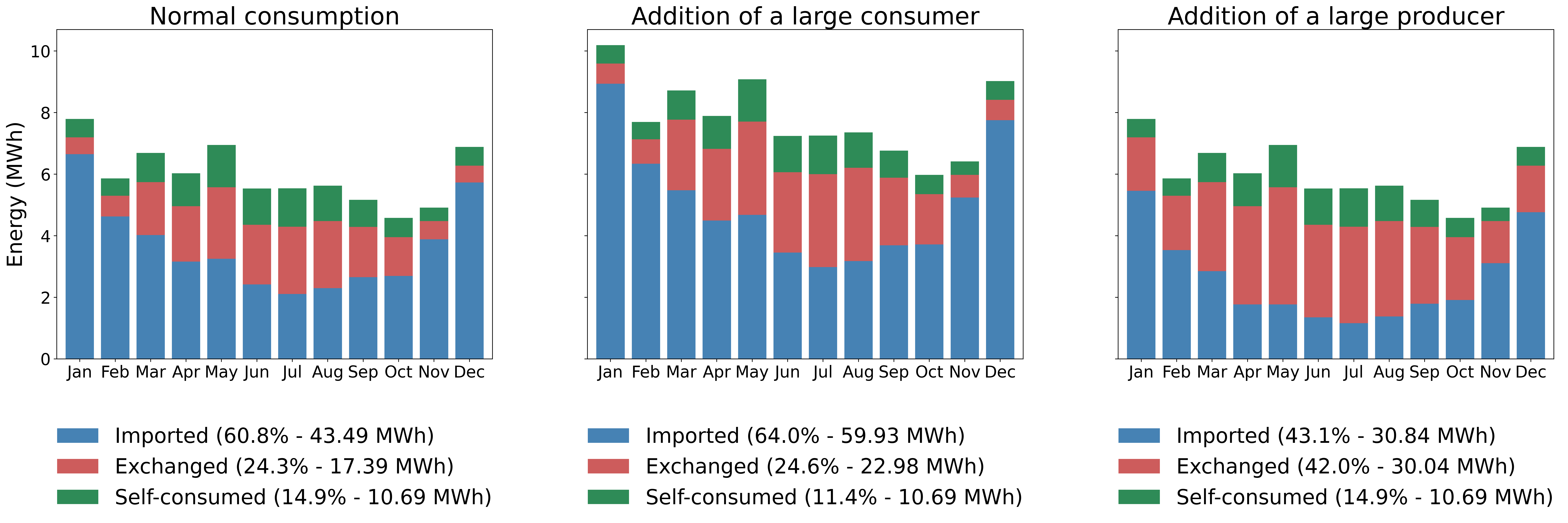}
    \caption{Monthly repartition of imported, exchanged, and self-consumed energy in CEL30 at 25\% PV penetration, comparing the baseline community with the addition of a large producer or a large consumer.}
    \label{fig:cel30_energy}
\end{figure}

The most illustrative results arise in CEL30 at 25\% PV penetration (Figure~\ref{fig:cel30_energy}). In the baseline community (left), only about 24\% of the total demand is covered by internal exchanges, with the majority supplied through imports (61\% – 43.5~MWh). When a large consumer is added (middle), redistribution increases only slightly to about 25\% of demand, while imports increase to 64\% (59.9~MWh), resulting in a less favorable balance. By contrast, the addition of a large PV producer (right) drastically increases redistribution within the CEL: 42\% of demand is then covered by internal exchanges (30.0~MWh), while imports drop to 43\% (30.8~MWh). 

\begin{figure}[H]
    \centering
    \includegraphics[width=1\textwidth]{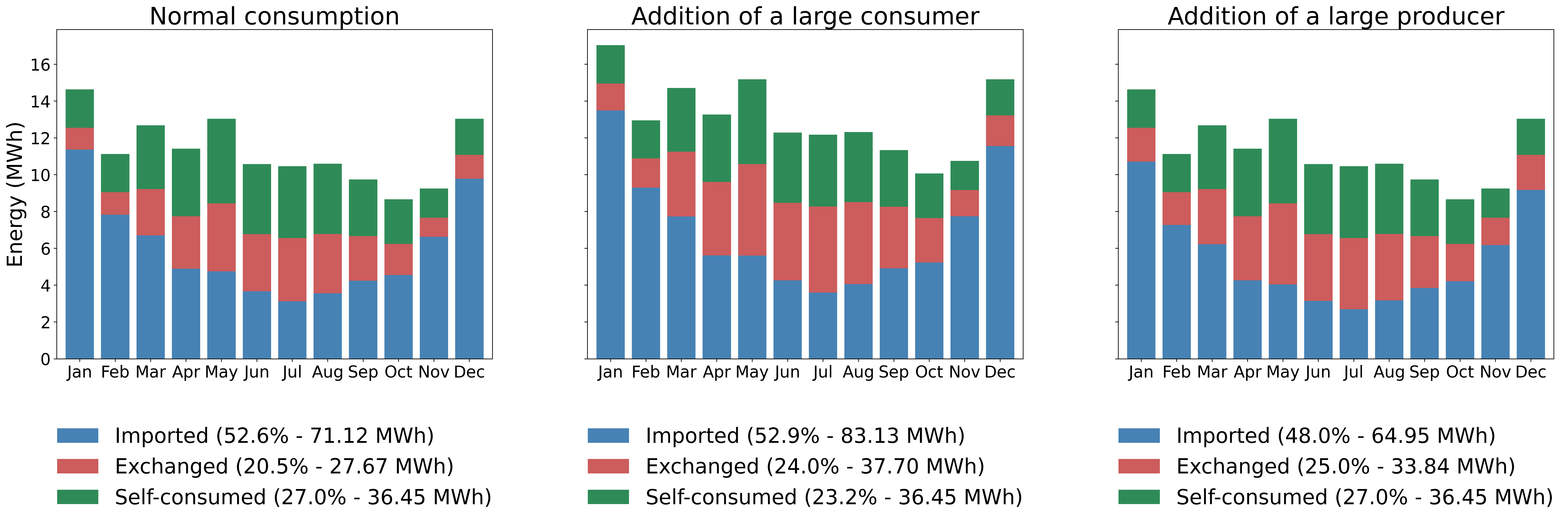}
    \caption{Monthly repartition of imported, exchanged, and self-consumed energy in CEL60 at 25\% PV penetration, comparing the baseline community with the addition of a large producer or a large consumer.}
    \label{fig:cel60_energy}
\end{figure}

This shows that introducing a producer-only member strongly reinforces cooperation at low penetration levels, as the other members can effectively absorb the surplus. Although this effect is shown here for CEL30, similar behavior is observed for larger communities as reported in Figure~\ref{fig:cel60_energy} and in~\ref{annex:eco_results} (Figure~\ref{fig:cel60_50_energy}).

\begin{figure}[H]
    \centering
    \includegraphics[width=0.9\textwidth]{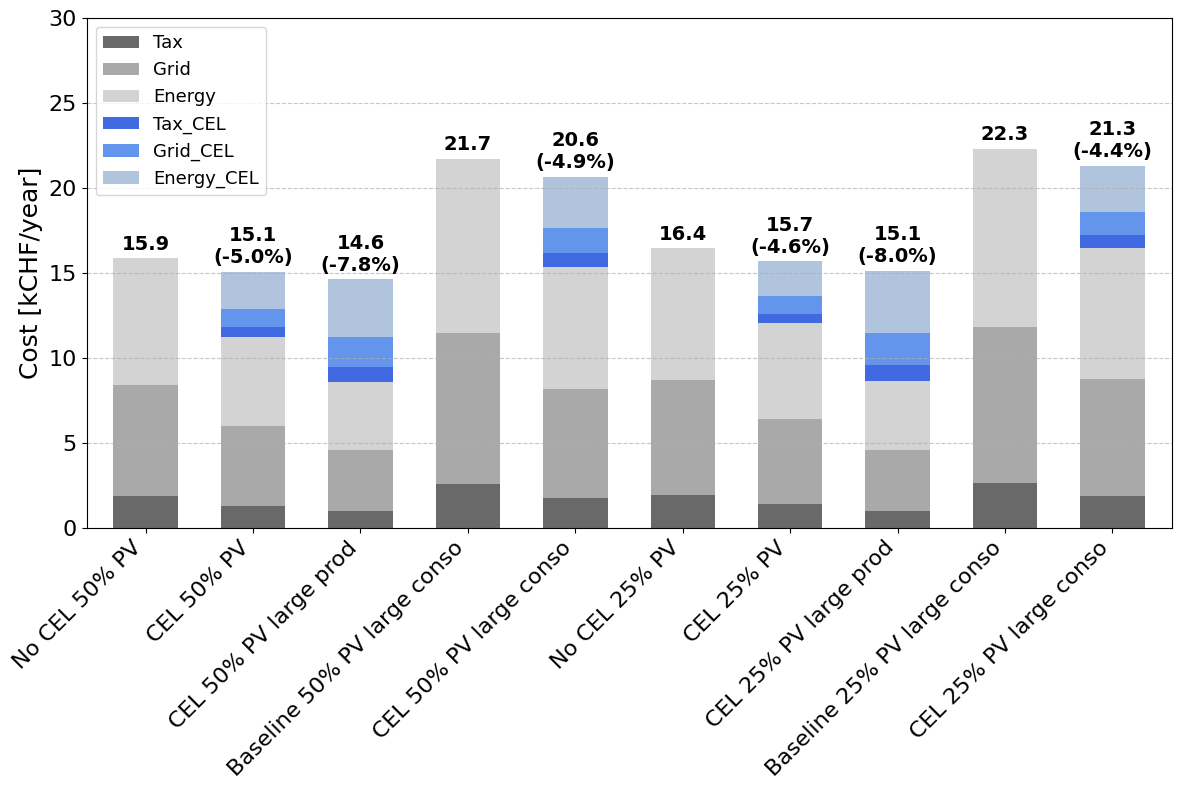}
    \caption{Annual electricity bill for CEL30, comparing the baseline community with the addition of a large producer or a large consumer for both 25 and 50\% PV penetration.}
    \label{fig:bill_cel30_large}
\end{figure}

Annual electricity bills confirm these dynamics (Figure~\ref{fig:bill_cel30_large}). With the addition of the large PV producer, the total bill decreases by 8\% compared to the non-CEL case, while the large consumer leads to savings of less than 5\%. The key driver is again the increase in redistribution: more local exchanges reduce distribution and energy charges, amplifying the benefits of community integration. 

Interestingly, the situation is reversed in CEL60 (Figure~\ref{fig:cel60_energy}). At both 25\% and 50\% PV penetration, the addition of a large consumer leads to more redistribution than the addition of a large producer. This is explained by the greater absorption capacity of the consumer in a larger and more diversified community: the surplus from existing PV installations can be more effectively reallocated to cover the demand of the consumer, while the contribution of a single large producer is diluted within the larger group. 

\begin{figure}[H]
    \centering
    \includegraphics[width=1\textwidth]{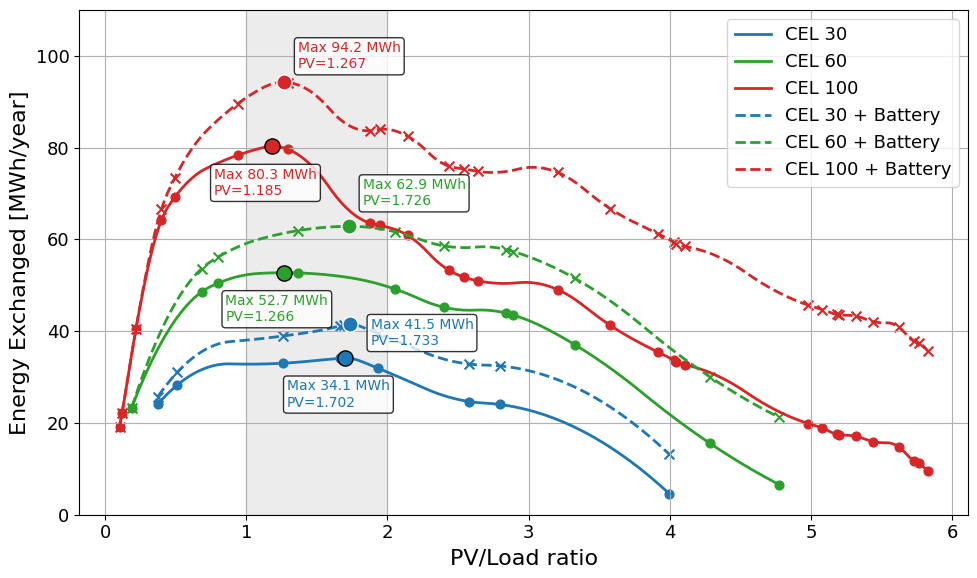}
    \caption{Energy exchanged within the community as a function of the ratio between PV generated energy (MWh/year) and annual load (MWh/year) for CEL30, CEL60, and CEL100, with and without a centralized battery.}
    \label{fig:exch_pvload}
\end{figure}

After observing these contrasting effects in CEL30 and CEL60 when adding a large producer versus a large consumer, the conditions under which one type of participant becomes more advantageous were analyzed. To generate Figure~\ref{fig:exch_pvload}, buildings equipped with PV systems were progressively added to different community sizes, and the annual internal energy exchanged within the respective CEL was recalculated after each addition. This procedure was applied to CEL30, CEL60, and CEL100, resulting in a denser set of configurations for larger communities and fewer configurations for smaller ones. For CEL30 and CEL60, the configurations shown correspond to communities located at the end of the line. However, several random combinations of buildings were also examined, confirming that the same overall trend is observed. The analysis was performed for both cases, with and without centralized storage.

The figure shows a consistent pattern: all community sizes (CEL30, CEL60, and CEL100, with or without centralized battery integration) reach maximum internal energy exchange when the PV-to-load ratio is between 1 and 2. The small “bumps” visible in the curves arise from the progressive addition of individual members with non-equivalent load and generation profiles, creating slight variations around the trend. Complementary analyses indicate that, for CEL30 and CEL60, the position of the optimum is not strongly affected by the order in which members are added. When the ratio falls below 1, meaning PV generation does not fully meet demand, the most advantageous strategy is to add a pure producer, increasing the surplus energy available to other members. In contrast, when the ratio exceeds 2, indicating significant overproduction relative to demand, adding a pure consumer is preferable because it helps absorb excess PV generation and reduces exports to the grid. The presence of a centralized battery shifts the point of maximum internal exchange to slightly higher PV-to-load ratios, as the battery can store excess generation and release it when needed, allowing for more PV capacity to be effectively integrated. In general, for the area studied, the optimal range for a CEL, in terms of maximizing internal exchanges while minimizing imports, is achieved by maintaining the PV-to-load ratio between 1 and 2 and strategically adding pure producers or consumers.

\subsubsection{Influence of Tariff Design and DSO Revenues}
\begin{table}[H]
\centering
\caption{Annual electricity bill and DSO revenue loss for CEL30 and CEL60 at PV penetration levels of 25\% and 50\% under double and dynamic tariffs, including percentage variations relative to the baselines. The baselines correspond to each PV penetration level without a CEL, while other scenarios represent a CEL with or without a centralized battery.}
\label{tab:bill_revenue_loss}
\renewcommand{\arraystretch}{1.2}
\begin{adjustbox}{max width=\textwidth}
\begin{tabular}{
l 
S[round-mode=places,round-precision=2]
S[table-format=+3.1,round-mode=places,round-precision=1]
S[round-mode=places,round-precision=2]
S[table-format=+3.1,round-mode=places,round-precision=1]
S[round-mode=places,round-precision=2]
S[table-format=+3.1,round-mode=places,round-precision=1]
S[round-mode=places,round-precision=2]
S[table-format=+3.1,round-mode=places,round-precision=1]
}
\toprule
 & \multicolumn{4}{c}{Double tariff} & \multicolumn{4}{c}{Dynamic tariff} \\
\cmidrule(lr){2-5} \cmidrule(lr){6-9}
Scenario 
& {Bill [kCHF]} & {Ref [\%]} 
& {Rev. Loss [kCHF]} & {Ref [\%]} 
& {Bill [kCHF]} & {Ref [\%]} 
& {Rev. Loss [kCHF]} & {Ref [\%]} \\
\midrule
\multicolumn{9}{l}{\textbf{CEL60}} \\
Baseline 50\% PV   & 28.726 & {--} & {--}   & {--} & 28.726 & {--} & {--}   & {--} \\
50\% PV            & 26.786 & -6.8 & 7.537  & +26.2 & 26.246 & -8.6 & 7.032  & +24.5 \\
50\% PV + Bat      & 26.033 & -9.4 & 10.281 & +35.8 & 25.214 & -12.2 & 10.019 & +34.9 \\
Baseline 25\% PV   & 32.867 & {--}  & {--}   & {--} & 32.867 & {--}  & {--}   & {--} \\
25\% PV            & 30.923 & -6.0 & 7.631  & +23.2 & 30.113 & -8.4 & 7.054  & +21.5 \\
25\% PV + Bat      & 30.501 & -7.2 & 9.134  & +27.8 & 29.298 & -10.9 & 8.762  & +26.7 \\
\midrule
\multicolumn{9}{l}{\textbf{CEL30}} \\
Baseline 50\% PV   & 20.612 & {--} & {--}   & {--} & 20.612 & {--} & {--}   & {--} \\
50\% PV            & 19.242 & -6.6 & 5.347  & +25.9 & 18.818 & -8.7 & 4.952  & +24.0 \\
50\% PV + Bat      & 18.822 & -8.7 & 6.878  & +33.4 & 18.214 & -11.6 & 6.607 & +32.1 \\
Baseline 25\% PV   & 22.667 & {--} & {--}   & {--} & 22.667 & {--} & {--}   & {--} \\
25\% PV            & 21.644 & -4.5 & 4.032  & +17.8 & 21.136 & -6.8 & 3.693  & +16.3 \\
25\% PV + Bat      & 21.470 & -5.3 & 4.651  & +20.5 & 20.753 & -8.5 & 4.403  & +19.4 \\
\bottomrule
\end{tabular}
\end{adjustbox}
\end{table}

As previously defined in Eq.~\ref{eq:revenue_loss}, the Revenue Loss metric presented in Table~\ref{tab:bill_revenue_loss} quantifies the financial impact on the DSO. It is calculated by subtracting the CEL-scenario bill (excluding the internal energy component, which is paid between members, not to the grid) from the total bill when there is no CEL. Therefore, this value represents the exact amount of revenue the DSO no longer collects due to the community's existence. The percentages shown in Table~\ref{tab:bill_revenue_loss} express the variation relative to each scenario’s baseline: negative values indicate reductions in annual bills, while the revenue loss percentages represent the share of the total baseline bill that the DSO no longer collects.

The first observation from Table~\ref{tab:bill_revenue_loss} is that the baseline scenarios result in identical bills regardless of the internal tariff scheme. This is expected, as baseline scenarios do not involve internal community exchanges; thus, all consumption is imported from the grid and billed under the standard external double tariff in both cases.

While forming a CEL results in savings for the members under both structures, the dynamic tariff provides a slight additional benefit over the double tariff, consistently leading to lower annual electricity bills. For example, in the CEL60 (50\% PV) scenario, the yearly bill under the double tariff is 26.79 kCHF, corresponding to a 6.8\% reduction relative to the baseline. Under the dynamic tariff, the bill further decreases to 26.25 kCHF, equivalent to an 8.6\% reduction, yielding roughly 2 percentage points of additional savings. This advantage becomes even more pronounced when batteries are integrated. In the same scenario with centralized storage, annual bills decrease from 26.03 kCHF (9.4\%) under the double tariff to 25.21 kCHF (12.2\%) under the dynamic tariff, widening the gap between the two schemes.

From the DSO perspective, the dynamic tariff also proves to be more favorable. In every simulated scenario, the revenue loss incurred by the grid operator is lower under the dynamic scheme than under the double tariff. Returning to the CEL60 (50\% PV) case, the DSO revenue loss falls from 7.54 kCHF (26.2\% of the baseline bill) under the double tariff to 7.03 kCHF (24.5\%). This indicates that the dynamic tariff provides a more balanced outcome, reducing member bills while limiting revenue losses for the DSO compared to the double tariff.

\section{Discussion} 
Local Electricity Communities influence participant economics and, when combined with distributed assets such as shared storage, can also affect local network operation. As decentralized generation and collective self-consumption continue to expand across Switzerland, understanding how CELs perform under different configurations becomes more important. The following discussion integrates economic and technical perspectives, contrasting the implications for rural community members and DSOs.

\subsection{The DSO perspective}
CELs do not inherently alter physical power flows just by their formation. Since energy exchanges occur over the public distribution network, the observed effects are solely attributable to the assets and their operation (such as PV penetration, location, and the centralized battery).

The centralized batteries used in the CEL30 and CEL60 cases have relatively small capacities. For instance, at 50\% PV penetration, the battery sizes are 30~kWh for CEL30 and 56~kWh for CEL60 while the corresponding PV capacities are 154 and 287~kWp, respectively. Regarding feed-in power, results follow the expected trend: higher PV penetration leads to increased feed-in peaks. Notably, for the cases with maximum PV penetration, the peak feed-in power remains unchanged relative to the baseline (462.1~kW for CEL60 and 263.5~kW for CEL30), despite the addition of a battery. This indicates that the economically optimal storage capacity is insufficient to absorb the significant generation surplus during peak PV generation. While a larger battery could technically reduce these feed-in peaks, it would not be economically justified, highlighting the trade-off between technical performance and economic viability.

An interesting effect emerges when examining grid imports. The increase in peak demand can be attributed to the fact that energy imported from the grid is billed under the double tariff, which interacts with the battery’s optimization strategy and shapes its charging behavior. To minimize costs, the battery tends to charge rapidly from the grid just before the price increases at 17:00. Since this charging coincides with building demand and in the absence of grid constraints, it produces a system peak that exceeds the peak observed without the centralized battery.

The location of the battery also influences line utilization: When placed near the transformer, some lines show slightly lower loading, whereas positioning it at the end of the line can increase utilization on specific branches. Voltage conditions follow a similar pattern. Over-voltage remains the main technical constraint at high PV penetration levels, occasionally reaching 1.10–1.12 p.u., while under-voltage issues are negligible. Overall, the centralized battery does not deliver the expected technical improvements, such as noticeable peak shaving, voltage stabilization, or line congestion relief, likely due to its limited capacity relative to total PV generation. A distributed storage configuration could potentially be more effective for both voltage control and congestion mitigation, enhancing technical resilience without compromising economic efficiency.

CELs also affect the economic balance between community members and DSOs. By lowering grid imports, CELs reduce the volume of electricity delivered through the distribution network, resulting in annual losses in distribution tariff revenues ranging from 4.03 to 10.28~kCHF, depending on community size, PV penetration, and the presence of batteries. Relative to the baseline bills, this corresponds to roughly 16--36\,\% of the DSO's annual revenue. In per-participant terms, this equates to approximately 397--541~CHF per participant per year (\(\approx\)33--45~CHF/month) in CEL60, and 448--764~CHF per participant per year (\(\approx\)37--64~CHF/month) in CEL30. These values indicate the potential reduction in DSO income associated with CEL operation, underlining that the economic gains achieved by community members translate into revenue losses for the DSO. Without regulatory adaptation, widespread CEL formation could therefore undermine cost recovery for grid operation.

The comparison between the double and dynamic tariffs shows that, while both yield similar technical outcomes, the dynamic model results in modestly lower annual bills for the community and a slight reduction in DSO revenue losses compared to the double tariff. Despite this economic advantage, local energy flows remain comparable between the two schemes. This is because the effective price differentials that matter for community members are modest: within the CEL, the spread is only about 9~cts/kWh for the double tariff (29.8--20.9~cts/kWh) and about 13~cts/kWh for the dynamic tariff (24.5--11.5~cts/kWh). Moreover, the limited capacity of the centralized battery constrains the ability to shift surplus PV generation, so using PV internally at noon is nearly cost-neutral compared to exporting it. These relatively small spreads and storage limitations provide limited incentive to shift demand or alter dispatch, meaning that the coincidence between PV generation and load remains the dominant driver of internal exchanges. As a result, internal exchanges remain nearly unchanged, indicating that under current tariff structures, physical generation-load matching outweighs the influence of tariff design.

\subsection{The CEL members' perspective}
For CEL participants, the economic benefits are clear. Collective operation lowers energy costs, improves local consumption, and redistributes expenses from recurring grid payments toward local investments in renewable generation.

PV penetration has a strong influence on CEL outcomes. Increasing PV capacity lowers the LCOE but also reduces the IRR, as higher upfront investments bring diminishing marginal savings. In CEL60, for instance, the LCOE decreases from 0.19 CHF/kWh at 25\% PV to 0.06 CHF/kWh at 80\%, while the IRR declines from 13.2\% to 9.6\%. A similar trend appears in CEL30, where the IRR drops from 14.8\% to about 11\% over the same range. Intermediate PV levels (around 25–50\%) therefore provide the best balance between affordability and investment performance. At 100\% PV penetration, the IRR falls further to 9.44\%. This occurs because when every member produces electricity simultaneously, internal exchanges sharply decline, leaving little opportunity for cooperation. As PV penetration increases, self-consumption becomes dominant, and the collective economic advantage of the CEL diminishes. In this sense, CELs cannot be considered as vehicles for maximizing PV installation alone, but rather as frameworks to optimize the trade-off between local generation and local consumption.

The lifetime cost analysis reinforces this balance: as PV capacity grows, expenditures gradually shift from operational to capital costs. In CEL60 and CEL30 with PV penetration higher than 50\%, operational revenues even outweigh maintenance expenses, resulting in negative OPEX values. This shift shows that members are gradually replacing external energy purchases with local assets, enhancing long-term self-sufficiency and reducing their dependence on retail electricity prices, which may fluctuate over time.

Community composition is another decisive factor. The inclusion of heterogeneous participants enhances internal balancing and improves the use of local generation. In smaller CELs such as CEL30 at lower PV penetration levels, adding a large PV producer significantly boosts internal exchanges. For instance, in CEL30 at 25\% PV, they rise from 24\% (17.4 MWh) to 42\% (30 MWh), while imports fall from 61\% (43.5 MWh) to 43\% (30.8 MWh). Conversely, including a large consumer in the same case produces only a minor effect, with exchanges increasing slightly to 24.6\% (23 MWh). At higher PV penetration levels, and particularly in larger communities such as CEL60, the trend reverses: adding a large consumer becomes more beneficial, as the group already has abundant generation and gains more from additional demand at 25\% PV, internal exchanges rise from 20\% (27.7 MWh) to 24\% (37.7 MWh) with a large consumer compared to 25\% (33.8 MWh) with a large producer. The same pattern holds in CEL100, where PV capacity is already abundant: the marginal gain from adding a large consumer is more relevant than from adding a large producer, even though the absolute effect is smaller than in CEL30 or CEL60 due to the already high diversity. Collectively, these observations align with the PV-to-load ratio analysis, indicating that CEL performance is maximized when local PV generation covers roughly 1–2 times the total community load. When the ratio is below 1, adding PV producers is the most effective strategy, whereas at ratios greater than 2, incorporating large consumers or flexible demand helps absorb excess generation and improves internal energy use.

The analysis of annual electricity bills confirmed that CEL formation systematically reduces participant costs. Total savings ranged between 5\% and 9\% compared to individual operation, with the largest reduction (9.4\%) observed in CEL60 at 50\% PV penetration, showing that larger and more diverse communities can achieve slightly higher collective gains. These savings are mainly driven by lower grid imports and by the 40\% reduction applied to the network usage component for internal exchanges. The battery further contributes to local exchanges, leading to a slight additional increase in savings. The demonstrated savings may not only attract new members to join a CEL but also encourage investment in larger shared batteries, as even small storage capacities already deliver measurable economic gains.

\subsection{Limitations and future research}
Several limitations constrain the generalizability of this study. Electric vehicles and heat pumps were excluded, meaning their potential to add flexible demand and additional load was not captured. Simulations were conducted at the building level without distinguishing between different household types or mixed-use buildings, which may influence both demand variability and self-consumption potential. In addition, the discretization of PV penetration levels is limited by the number of participants. Since PV installation is defined at the building level, certain intermediate penetration values cannot be represented precisely. As an example, in CEL30 (9 buildings), both 25\% and 30\% PV penetration correspond to two PV-equipped buildings, while 35\% and 40\% correspond to three. This limits the resolution of the sensitivity analysis, though the effect diminishes in larger communities such as CEL60 (19 buildings). Tariff-switching behavior (for instance, moving away from H4 tariffs after joining a CEL) was also not considered, even though it could affect member costs and incentives.

Future research should extend this analysis to include electric vehicles, heat pumps, and other flexible assets to better quantify the role of demand-side management within CELs. Applying the framework to urban areas would also be valuable, as higher network density, mixed-use profiles, and multi-apartment buildings could significantly alter the balance between local exchanges and grid reliance. In addition, future studies should investigate the effects of decentralized battery systems, assessing how distributing storage across individual buildings rather than relying on a single central unit influences overall community performance. Further work should also explore different repartition key designs to evaluate how alternative cost and benefit allocation mechanisms affect member incentives and fairness. Finally, research should aim to identify the optimal CEL configuration, specifically the best match between producers and consumers to maximize economic benefits while ensuring equitable cost sharing among participants.

\section{Conclusion}
This study evaluated the techno-economic performance of CELs in a rural Swiss context, assessing the influence of PV penetration, community composition, shared storage, and tariff design. The findings confirm that CELs can provide substantial economic benefits and support the integration of renewable energy when properly designed to balance local generation and demand. Their performance depends primarily on how assets are sized, located, and combined, rather than on the CEL framework itself.

From the DSO perspective, CELs have limited influence on grid operation, as energy exchanges still occur through the public distribution network. The centralized batteries used in this study deliver clear economic value, but their technical impact remains modest due to their small capacity relative to installed PV. While scaling up centralized storage capacity can reduce transformer peak load, it risks increasing line loading. Distributed storage could extend this benefit to voltage regulation and peak load reduction. Importantly, tariff structures can influence grid stress: they may help mitigate it, but in some cases, they can also exacerbate it. For example, the double tariff encourages the centralized battery to charge aggressively before peak hours, creating a new peak load. Economically, for the studied cases, CELs lower grid imports by 27–46\%, which in turn reduces DSO tariff revenues by roughly 17–36\% of annual income, underscoring the need to adapt cost-recovery mechanisms. While CELs do not fundamentally alter grid operation, they provide a valuable framework within which flexibility can emerge, provided that suitable incentives and market signals are introduced. Current tariff reductions applied mainly to distribution tariffs already deliver substantial savings for participants, but stimulating investments in larger or more responsive storage systems will be key to unlocking CELs’ full technical potential and transforming them into active contributors to grid stability and local flexibility.

From the members’ perspective, CELs enable participants to reduce energy costs and increase their autonomy by relying more on local renewable generation. They should not be seen as mechanisms to maximize PV capacity, but rather as frameworks to balance local production and consumption efficiently. Increasing PV capacity reduces the LCOE but also lowers investment returns, with IRRs declining from around 13–15\% at moderate levels to roughly 9\% at full PV coverage. Despite this drop, such returns remain attractive in Switzerland, proving that CELs can remain profitable even under high PV penetration. Community performance depends more on the diversity and complementarity of members than on size alone: for CELs with PV-to-load ratios below 1, adding large producers enhances local exchanges, while for ratios above 2, incorporating large consumers that absorb excess generation improves internal energy use. The analysis showed that even small shared batteries improve local exchanges, reducing annual electricity bills by up to 9.4\% compared to individual operation without a CEL, thus improving overall economic performance. Greater value could be achieved by fostering diverse member profiles and integrating flexible assets such as electric vehicles, heat pumps, or demand-side management, allowing communities to dynamically balance generation and consumption, which would not only reinforce their contribution to the energy transition but also provide benefits to DSOs by reducing network stress and limiting the need for costly grid reinforcements.

\bibliographystyle{ieeetr}
\bibliography{references}

\appendix
\section{Glossary of optimization model parameters}
\label{annex:parameters}

\begin{table}[H]
\centering
\begin{tabular}{ll}
\toprule
Symbol & Definition \\\midrule
\(TOTEX\)        & Total cost of ownership over the system lifetime \\
\(OPEX\)         & Operating expenditures (grid exchanges, battery operation, PV maintenance) \\
\(CAPEX\)        & Capital expenditures (PV and battery investments) \\
\(R\)            & Annuity factor accounting for discount rate and lifetime \\
\(L\)            & System lifetime (years) \\
\(L^{bat}\)      & Battery lifetime (years) \\
\(r\)            & Discount rate \\
\(OX_{ge}\)      & Grid exchange costs \\
\(OX_{bo}\)      & Battery operation costs \\
\(OX_{pm}\)      & PV maintenance costs \\
\(P_t^{IMP}\)    & Imported power at time step \(t\) \\
\(P_t^{IMP}\)   & Corrective term for imported power (if applicable) \\
\(P_t^{EXP}\)    & Exported power at time step \(t\) \\
\(TS_t\)         & Duration of time step \(t\) \\
\(CX_{pv}\)      & PV investment cost \\
\(CX_{bat}\)     & Battery investment cost \\
\(\mu_i^{MOD}\)  & Number of PV modules of type \(i\) \\
\(P_{nom,i}^{MOD}\) & Nominal power of PV module \(i\) \\
\(C^{MOD}\)      & Specific cost per PV module \\
\(\beta^W\)      & Indicator variable for PV wiring and balance-of-system costs \\
\(C^{FW}\)       & Fixed PV wiring and installation cost \\
\(E_{cap,j}^{BAT}\) & Capacity of battery unit \(j\) \\
\(C^{BAT}_{sp}\) & Specific cost of battery capacity (CHF/kWh) \\
\(\beta^{BAT}\)  & Indicator variable for battery fixed costs \\
\(C^{BAT}_{fix}\)& Fixed battery installation cost \\
\(C^{BAT}_d\)    & Specific cost of battery discharge (CHF/kWh) \\
\(C^{BAT}_c\)    & Specific cost of battery charging (CHF/kWh) \\
\(\gamma^M\)     & Maintenance cost factor (fraction of PV CAPEX per year) \\
\bottomrule
\end{tabular}
\caption{Definition of parameters used in the optimization model.}
\label{tab:parameters_model}
\end{table}

\newpage
\section{Tariff Structures}
\label{annex:tariff_breakdown}

This appendix details the pricing components used in the study. External grid imports are always billed under the Double Tariff. For internal community exchanges (CEL), two distinct models are evaluated: the Double Tariff (with 40\% grid usage reduction) and the Dynamic Tariff (irradiance-based).

\subsection{Double Tariff}
Table~\ref{tab:tariff_breakdown} details the 2025 Double Tariff components (profile ``Energie Suisse''), contrasting the standard rates for external imports against the reduced rates for internal exchanges.

\begin{table}[H]
\centering
\caption{Breakdown of the 2025 Double Tariff components. Internal exchanges benefit from a 40\% reduction on grid usage fees.}
\label{tab:tariff_breakdown}
\renewcommand{\arraystretch}{1.2}
\setlength{\tabcolsep}{4pt}
\begin{tabular}{lccccc}
\toprule
& \multicolumn{2}{c}{\textbf{Standard (DSO)}} & \multicolumn{2}{c}{\textbf{Internal (CEL)}} & \\
\cmidrule(lr){2-3} \cmidrule(lr){4-5}
\textbf{Component} & \textbf{HP} & \textbf{HC} & \textbf{HP} & \textbf{HC} & \textbf{Unit} \\
\midrule
\multicolumn{6}{l}{\textbf{Energy Supply} \textit{(Time-of-Use)}} \\
Energy Price & 16.68 & 11.81 & 16.68 & 11.81 & cts./kWh \\
\midrule
\multicolumn{6}{l}{\textbf{Grid Usage} \textit{(Fixed - 40\% reduction for CEL)}} \\
Regional Grid & 14.34 & 8.43 & 8.60 & 5.06 & cts./kWh \\
National Grid (Swissgrid) & 2.32 & 1.48 & 1.39 & 0.89 & cts./kWh \\
\midrule
\multicolumn{6}{l}{\textbf{Public Taxes} \textit{(Fixed)}} \\
Federal Taxes & \multicolumn{4}{c}{2.30} & cts./kWh \\
Winter Electricity Reserve & \multicolumn{4}{c}{0.23} & cts./kWh \\
Cantonal Tax & \multicolumn{4}{c}{0.60} & cts./kWh \\
Cantonal Emolument & \multicolumn{4}{c}{0.02} & cts./kWh \\
\midrule
\textbf{Total Variable Cost} & \textbf{36.49} & \textbf{24.87} & \textbf{29.82} & \textbf{20.91} & \textbf{cts./kWh} \\
\bottomrule
\end{tabular}
\end{table}

\subsection{Dynamic Tariff}
In this model, internal exchanges follow a pricing logic driven by solar availability. The Grid Usage and Public Taxes are treated as fixed constants, while the \textbf{Energy Supply} component fluctuates to match the total price derived from irradiance levels.

\paragraph{Component Calculation}
\begin{itemize}
    \item \textbf{Grid Usage:} Fixed at 7.39 ct./kWh (derived from the Simple Tariff with the 40\% reduction applied).
    \item \textbf{Public Taxes:} Fixed at 3.15 ct./kWh.
    \item \textbf{Energy Supply:} Variable. Calculated as the residual value: $P_{energy}(t) = P_{total}(t) - P_{fixed}$.
\end{itemize}

\begin{table}[H]
\centering
\caption{Breakdown of the Internal Dynamic Tariff. The Grid and Tax components are fixed, while the Energy Supply component varies.}
\label{tab:dynamic_tariff_breakdown}
\renewcommand{\arraystretch}{1.2}
\begin{tabular}{lcc}
\toprule
\textbf{Component} & \textbf{Value / Range} & \textbf{Unit} \\
\midrule
\multicolumn{3}{l}{\textbf{Fixed Components}} \\
Grid Usage (Regional + National) \textsuperscript{*} & 7.39 & ct./kWh \\
Public Taxes & 3.15 & ct./kWh \\
\textbf{Total Fixed Costs} & \textbf{10.54} & \textbf{ct./kWh} \\
\midrule
\multicolumn{3}{l}{\textbf{Variable Component}} \\
\textbf{Energy Supply} & \textbf{0.96 -- 13.98} & \textbf{ct./kWh} \\
\midrule
\textbf{Total Dynamic Price (CEL)} & \textbf{11.50 -- 24.52} & \textbf{ct./kWh} \\
\bottomrule
\multicolumn{3}{l}{\footnotesize \textsuperscript{*} Includes the 40\% regulatory reduction on grid fees.} \\
\end{tabular}
\end{table}

\paragraph{Minimum Price Logic}
Under maximum irradiance conditions, the total dynamic price drops to its minimum floor of 11.50 ct./kWh. At this point, the Energy Supply component reaches its minimum value ($11.50 - 10.54 = 0.96$ ct./kWh), ensuring that all fixed grid and tax obligations remain covered.

\newpage
\section{Lines data}
\label{annex:line_parameters}
\begin{table}[H]
\centering
\caption{Line characteristics for the rural LV network, showing segment identifiers, lengths, and impedance values.}
\begin{tabular}{lccc}
\hline
\textbf{Nom tronçon} & \textbf{Longueur (m)} & \textbf{R1 ($\Omega$)} & \textbf{X1 ($\Omega$)} \\
\hline
401663523 & 69.0 & 0.034 & 0.006 \\
401663529 & 24.0 & 0.012 & 0.002 \\
43928166 & 35.0 & 0.008 & 0.003 \\
43923879 & 8.0 & 0.002 & 0.001 \\
43970837 & 9.0 & 0.002 & 0.001 \\
43970839 & 75.0 & 0.007 & 0.005 \\
43970838 & 20.0 & 0.005 & 0.002 \\
43817110 & 15.0 & 0.004 & 0.001 \\
43970830 & 33.0 & 0.009 & 0.003 \\
43818057 & 37.0 & 0.010 & 0.003 \\
43850647 & 3.0 & 0.003 & 0.000 \\
43974044 & 22.0 & 0.020 & 0.002 \\
43970841 & 35.0 & 0.051 & 0.003 \\
43970840 & 9.0 & 0.013 & 0.001 \\
43970820 & 48.0 & 0.013 & 0.004 \\
43970816 & 10.0 & 0.003 & 0.001 \\
43897486 & 40.0 & 0.021 & 0.013 \\
43821648 & 22.0 & 0.035 & 0.008 \\
43854952 & 36.0 & 0.048 & 0.003 \\
43994083 & 25.0 & 0.037 & 0.002 \\
43970821 & 2.0 & 0.003 & 0.000 \\
43970819 & 17.0 & 0.023 & 0.002 \\
109249426 & 13.0 & 0.003 & 0.001 \\
109249452 & 43.0 & 0.063 & 0.004 \\
43985272 & 63.0 & 0.017 & 0.005 \\
151640501 & 32.0 & 0.008 & 0.002 \\
43829950 & 40.0 & 0.011 & 0.003 \\
151640973 & 40.0 & 0.037 & 0.003 \\
43943613 & 4.0 & 0.006 & 0.000 \\
43909489 & 16.0 & 0.023 & 0.001 \\
43985275 & 80.0 & 0.117 & 0.007 \\
43985271 & 22.0 & 0.029 & 0.002 \\
\hline
\end{tabular}
\label{tab:line_parameters}
\end{table}

\begin{landscape}
\section{Technical results}
\label{annex:tech_results}
\setlength{\tabcolsep}{3pt}
\begin{table}[H]
\centering
\scriptsize
\begin{tabular}{ll
  >{\centering\arraybackslash}p{1.05cm}
  >{\centering\arraybackslash}p{1.05cm}
  >{\centering\arraybackslash}p{1.05cm}
  >{\centering\arraybackslash}p{1.05cm}
  >{\centering\arraybackslash}p{1.05cm}
  >{\centering\arraybackslash}p{1.05cm}
  >{\centering\arraybackslash}p{1.05cm}
  >{\centering\arraybackslash}p{1.05cm}
  >{\centering\arraybackslash}p{1.05cm}
  >{\centering\arraybackslash}p{1.05cm}
  >{\centering\arraybackslash}p{1.05cm}
  >{\centering\arraybackslash}p{1.05cm}
  >{\centering\arraybackslash}p{1.05cm}
}
\toprule
Network & Line & Baseline & CEL100 bat up & CEL100 bat down & CEL100 50PV & CEL100 25PV & Baseline CEL60 & CEL60 100PV & CEL60 50PV & CEL60 25PV & Baseline CEL30 & CEL30 100PV & CEL30 50PV & CEL30 25PV \\
\midrule
Double Tariff & 40 & 105.13 & 103.84 & 103.84 & 64.31 & 31.05 & 23.67 & 23.67 & 22.54 & 10.51 & 14.16 & 14.16 & 14.20 & 14.22 \\
Dynamic tariff & 40 & 105.13 & 103.84 & 103.84 & 64.31 & 31.05 & 23.67 & 23.67 & 22.54 & 10.51 & 14.16 & 14.16 & 14.20 & 14.22 \\
Double Tariff & 46 & 97.01 & 96.79 & 96.79 & 34.24 & 34.51 & 34.37 & 34.37 & 34.51 & 34.64 & 34.56 & 34.56 & 34.65 & 34.71 \\
Dynamic tariff & 46 & 97.01 & 96.79 & 96.79 & 34.24 & 34.51 & 34.37 & 34.37 & 34.51 & 34.64 & 34.56 & 34.56 & 34.65 & 34.71 \\
Double Tariff & 39 & 81.40 & 78.95 & 78.95 & 38.42 & 29.33 & 79.59 & 79.59 & 24.96 & 18.61 & 65.28 & 65.28 & 26.82 & 14.38 \\
Dynamic tariff & 39 & 81.40 & 78.95 & 78.95 & 34.49 & 27.39 & 79.59 & 79.59 & 24.96 & 18.61 & 65.28 & 65.28 & 26.82 & 14.38 \\
Double Tariff & 9  & 77.61 & 77.43 & 77.43 & 27.39 & 27.60 & 27.49 & 27.49 & 27.60 & 27.71 & 27.65 & 27.65 & 27.72 & 27.76 \\
Dynamic tariff & 9  & 77.61 & 77.43 & 77.43 & 27.39 & 27.60 & 27.49 & 27.49 & 27.60 & 27.71 & 27.65 & 27.65 & 27.72 & 27.76 \\
Double Tariff & 1  & 51.14 & 50.64 & 50.64 & 50.26 & 24.30 & 14.84 & 14.84 & 13.94 & 6.59 & 6.59 & 6.59 & 6.59 & 6.59 \\
Dynamic tariff & 1  & 51.14 & 50.64 & 50.64 & 50.26 & 24.30 & 14.84 & 14.84 & 13.94 & 6.59 & 6.59 & 6.59 & 6.59 & 6.59 \\
Double Tariff & 28 & 50.61 & 49.45 & 49.45 & 50.00 & 11.03 & 11.03 & 11.03 & 11.03 & 11.03 & 11.03 & 11.03 & 11.03 & 11.03 \\
Dynamic tariff & 28 & 50.61 & 49.45 & 49.45 & 50.00 & 11.03 & 11.03 & 11.03 & 11.03 & 11.03 & 11.03 & 11.03 & 11.03 & 11.03 \\
\bottomrule
\end{tabular}
\caption{Maximum percentage loading for the 6 most loaded lines in baseline across all scenarios.}
\label{tab:most_loaded_max}
\end{table}

\setlength{\tabcolsep}{3pt}
\begin{table}[H]
\centering
\scriptsize

\begin{tabular}{ll
  >{\centering\arraybackslash}p{1.0cm}
  >{\centering\arraybackslash}p{1.0cm}
  >{\centering\arraybackslash}p{1.0cm}
  >{\centering\arraybackslash}p{1.0cm}
  >{\centering\arraybackslash}p{1.0cm}
  >{\centering\arraybackslash}p{1.0cm}
  >{\centering\arraybackslash}p{1.0cm}
  >{\centering\arraybackslash}p{1.0cm}
  >{\centering\arraybackslash}p{1.0cm}
  >{\centering\arraybackslash}p{1.0cm}
  >{\centering\arraybackslash}p{1.0cm}
  >{\centering\arraybackslash}p{1.0cm}
  >{\centering\arraybackslash}p{1.0cm}
}
\toprule
Network & Line & Baseline & CEL100 bat up & CEL100 bat down & CEL100 50PV & CEL100 25PV &
Baseline CEL60 & CEL60 max PV & CEL60 50PV & CEL60 25PV &
Baseline CEL30 & CEL30 max PV & CEL30 50PV & CEL30 25PV \\
\midrule
Double Tariff & 40 & 41.35 & 43.31 & 41.61 & 24.86 & 12.04 & 11.94 & 11.98 & 9.92 & 4.04 & 5.92 & 5.87 & 3.51 & 2.97 \\
Dynamic Tariff & 40 & 41.35 & 43.31 & 41.61 & 24.84 & 12.03 & 11.94 & 11.98 & 9.92 & 4.04 & 5.92 & 5.87 & 3.51 & 2.97 \\
Double Tariff & 46 & 39.43 & 35.58 & 39.63 & 10.75 & 5.83 & 9.48 & 9.54 & 6.50 & 4.80 & 6.41 & 6.37 & 4.09 & 3.67 \\
Dynamic Tariff & 46 & 39.43 & 35.58 & 39.63 & 10.75 & 5.82 & 9.48 & 9.54 & 6.50 & 4.80 & 6.41 & 6.36 & 4.09 & 3.67 \\
Double Tariff & 39 & 34.95 & 35.11 & 36.41 & 16.20 & 6.02 & 28.46 & 29.57 & 8.79 & 4.52 & 19.94 & 20.71 & 5.75 & 3.42 \\
Dynamic Tariff & 39 & 34.95 & 35.11 & 36.41 & 16.15 & 6.03 & 28.46 & 29.45 & 8.79 & 4.52 & 19.94 & 20.67 & 5.75 & 3.42 \\
Double Tariff & 9  & 34.32 & 32.24 & 34.42 & 11.09 & 5.64 & 9.78 & 9.83 & 6.56 & 4.53 & 6.45 & 6.37 & 3.87 & 3.28 \\
Dynamic Tariff & 9  & 34.32 & 32.24 & 34.42 & 11.03 & 5.63 & 9.78 & 9.83 & 6.56 & 4.53 & 6.45 & 6.37 & 3.87 & 3.28 \\
Double Tariff & 1  & 25.45 & 25.78 & 25.69 & 20.94 & 10.42 & 9.24 & 9.28 & 7.62 & 3.80 & 5.70 & 5.70 & 3.10 & 2.54 \\
Dynamic Tariff & 1  & 25.45 & 25.78 & 25.69 & 20.93 & 10.41 & 9.24 & 9.28 & 7.61 & 3.80 & 5.70 & 5.70 & 3.10 & 2.54 \\
Double Tariff & 28 & 24.95 & 25.21 & 25.13 & 20.83 & 5.70 & 8.41 & 8.50 & 6.04 & 4.77 & 6.01 & 5.95 & 4.04 & 3.27 \\
Dynamic Tariff & 28 & 24.95 & 25.21 & 25.13 & 20.80 & 5.70 & 8.41 & 8.50 & 6.04 & 4.77 & 6.01 & 5.95 & 4.04 & 3.27 \\
\bottomrule
\end{tabular}
\caption{Median percentage loading for the 6 most loaded lines in baseline across all scenarios.}
\label{tab:most_loaded_median}
\end{table}
\end{landscape}

\begin{table}[H]
\centering
\caption{Maximum feed-in and drawn power values at the transformer level for different CEL scenarios in kW, expressed as absolute values and as percentages relative to the baseline of each CEL size.}
\label{tab:cel_scenarios}
\renewcommand{\arraystretch}{1.2}
\begin{adjustbox}{max width=\textwidth}
\begin{tabular}{l *{8}{S[table-format=4.2]}}
\toprule
 & \multicolumn{4}{c}{DT\_suisse} & \multicolumn{4}{c}{irr\_15} \\
\cmidrule(lr){2-5} \cmidrule(lr){6-9}
Name 
& {Feed-in power} & {Reference \%} 
& {Power drawn} & {Reference \%} 
& {Feed-in power} & {Reference \%} 
& {Power drawn} & {Reference \%} \\
\midrule
Baseline            
& 934.27 & 100.0 & 81.48 & 100.0
& 934.27 & 100.0 & 81.48 & 100.0 \\

CEL100 max PV down  
& 928.06 & 99.3 & 154.73 & 189.9
& 928.06 & 99.3 & 154.73 & 189.9 \\

CEL100 max PV up    
& 928.06 & 99.3 & 150.79 & 185.1
& 928.06 & 99.3 & 150.79 & 185.1 \\

CEL100 50PV         
& 544.49 & 58.3 & 174.07 & 213.7
& 544.49 & 58.3 & 81.48 & 100.0 \\

CEL100 25PV         
& 198.62 & 21.3 & 136.25 & 167.3
& 198.62 & 21.3 & 81.48 & 100.0 \\
\midrule
Baseline CEL60      
& 462.08 & 100.0 & 81.48 & 100.0
& 462.08 & 100.0 & 81.48 & 100.0 \\

CEL60 max PV        
& 462.08 & 100.0 & 119.56 & 146.8
& 462.08 & 100.0 & 119.56 & 146.8 \\

CEL60 50PV          
& 213.55 & 46.2 & 118.02 & 144.9
& 213.55 & 46.2 & 118.02 & 144.9 \\

CEL60 25PV          
& 57.96 & 12.6 & 91.69 & 112.5
& 57.96 & 12.6 & 91.69 & 112.5 \\
\midrule
Baseline CEL30      
& 263.48 & 100.0 & 81.48 & 100.0
& 263.48 & 100.0 & 81.48 & 100.0 \\

CEL30 max PV        
& 263.48 & 100.0 & 92.24 & 113.2
& 263.48 & 100.0 & 92.24 & 113.2 \\

CEL30 50PV          
& 97.69 & 37.1 & 88.50 & 108.6
& 97.69 & 37.1 & 88.50 & 108.6 \\

CEL30 25PV          
& 20.66 & 7.8 & 81.48 & 100.0
& 20.66 & 7.8 & 81.48 & 100.0 \\
\bottomrule
\end{tabular}
\end{adjustbox}
\end{table}

\begin{figure}[H]
\centering
\includegraphics[width=1\textwidth]{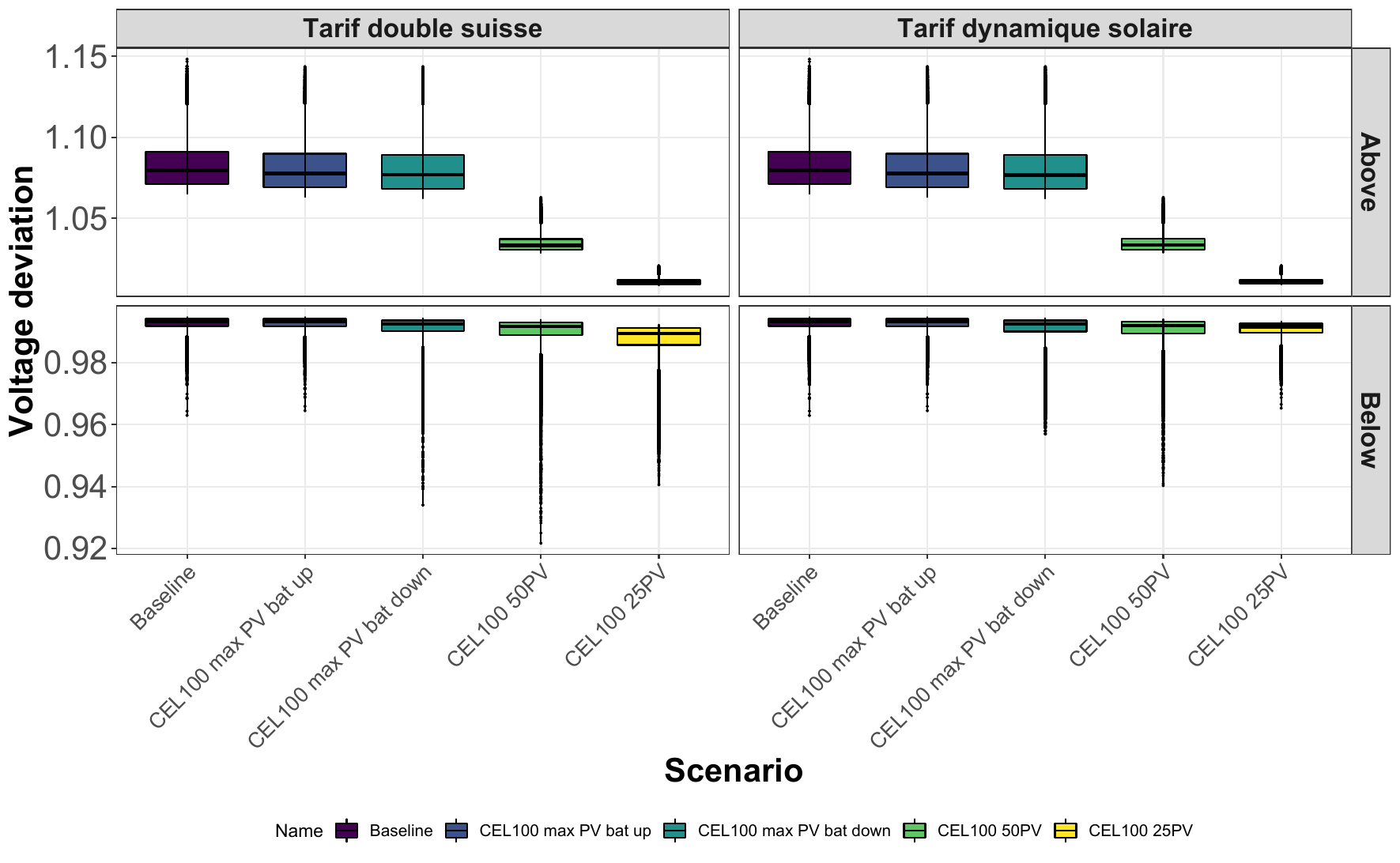}
\caption{Voltage deviation distribution (95th percentile) for CEL100.}
\label{fig:voltage_cel100}
\end{figure}

\section{Building data}
\label{annex:building_data}
\begin{table}[H]
\centering
\scriptsize
\begin{tabular}{c c c c c c}
\hline
Building & Load [MWh/year] & PV [MWh/year] & PV installed [kWp] & Self-consumed [MWh] & Self-Sufficiency [\%] \\
\hline
1  & 6.6  & 115.5 & 113.4 & 3.7 & 55.9 \\
2  & 7.0  & 41.3  & 40.3  & 2.8 & 40.4 \\
3  & 22.0 & 178.9 & 172.3 & 14.0 & 63.6 \\
4  & 4.2  & 12.8  & 12.9  & 1.8 & 42.2 \\
5  & 11.2 & 34.5  & 33.7  & 5.8 & 51.8 \\
6  & 2.5  & 36.9  & 36.2  & 1.4 & 54.4 \\
7  & 1.7  & 12.5  & 12.3  & 0.7 & 39.4 \\
8  & 4.1  & 22.6  & 22.1  & 2.0 & 48.9 \\
9  & 4.2  & 21.1  & 20.8  & 2.0 & 47.8 \\
10 & 2.8  & 28.7  & 29.3  & 1.1 & 39.9 \\
11 & 31.6 & 110.8 & 107.1 & 21.0 & 66.4 \\
12 & 6.5  & 22.4  & 22.7  & 3.2 & 49.1 \\
13 & 7.0  & 39.9  & 39.4  & 3.4 & 48.5 \\
14 & 15.3 & 69.4  & 71.5  & 6.6 & 42.9 \\
15 & 4.1  & 20.0  & 19.5  & 1.1 & 26.3 \\
16 & 3.9  & 14.3  & 13.9  & 1.0 & 24.5 \\
17 & 1.8  & 7.6   & 7.2   & 0.6 & 30.8 \\
18 & 0.9  & 3.6   & 3.2   & 0.4 & 42.7 \\
19 & 0.6  & 0.0   & 0.0   & 0.0 & 0.0  \\
20 & 16.6 & 59.4  & 60.8  & 9.1 & 54.5 \\
21 & 5.4  & 21.7  & 23.0  & 1.5 & 28.1 \\
22 & 4.8  & 22.0  & 22.1  & 2.0 & 40.9 \\
23 & 1.7  & 6.3   & 6.0   & 0.5 & 30.1 \\
24 & 3.5  & 70.5  & 69.6  & 1.6 & 45.7 \\
25 & 0.2  & 2.6   & 2.5   & 0.1 & 35.4 \\
26 & 11.4 & 34.0  & 32.5  & 4.9 & 42.9 \\
27 & 3.4  & 19.1  & 18.6  & 1.5 & 43.7 \\
28 & 2.5  & 19.7  & 19.2  & 1.2 & 46.3 \\
29 & 3.9  & 18.5  & 17.6  & 1.7 & 43.0 \\
30 & 3.9  & 25.4  & 25.5  & 1.3 & 32.8 \\
31 & 7.0  & 72.8  & 71.5  & 5.2 & 73.8 \\
32 & 1.4  & 24.3  & 23.6  & 0.5 & 36.3 \\
\hline
\end{tabular}
\caption{Load, PV generation, installed PV capacity, self-consumed energy and self-sufficiency for each building.}
\label{tab:building_data}
\end{table}

\section{Economical results}
\label{annex:eco_results}
\begin{table}[H]
\centering
\caption{Techno-economic results for different CEL cases}
\resizebox{\linewidth}{!}{%
\begin{tabular}{ll
S[round-mode=places,round-precision=4] 
S[round-mode=places,round-precision=1] 
S[round-mode=places,round-precision=1] 
S[round-mode=places,round-precision=1]}
\hline
\textbf{Case} & \textbf{Scenario} & \textbf{LCOE (CHF/kWh)} & \textbf{IRR (\%)} & \textbf{DPP (years)} & \textbf{Profit (kCHF)} \\
\hline
\multirow{4}{*}{CEL100} 
 & 50\% PV     & 0.0538 & 10.3067995356254 & 10.4969199178644  & 772.886 \\
 & 50\% PV + Bat    & 0.0453 & 10.4198861047641 & 10.4969199178644  & 803.337 \\
 & 25\% PV     & 0.1702 & 11.0759045935588 & 10.0862422997946  & 354.39 \\
 & 25\% PV + Bat    & 0.1645 & 11.2266354164689 & 10.1629021218343  & 374.918 \\
\hline
\multirow{4}{*}{CEL60} 
 & 50\% PV     & 0.1017 & 10.9278677857822 & 10.1629021218343 & 375.79 \\
 & 50\% PV + Bat    & 0.0943 & 11.047694441302 & 10.2477754962354 & 392.602 \\
 & 25\% PV     & 0.1868 & 13.2326268451571 & 8.49828884325804  & 182.56 \\
 & 25\% PV + Bat    & 0.1826 & 13.313824924638 & 8.33127994524298  & 192.163 \\
\hline
\multirow{4}{*}{CEL30} 
 & 50\% PV     & 0.1198 & 12.5935808324359 & 9.00205338809035  & 239.763 \\
 & 50\% PV + Bat    & 0.1136 & 12.7022014554216 & 8.58316221765913  & 249.851 \\
 & 25\% PV     & 0.2038 & 14.8260115760244 & 7.58110882956878  & 103.001 \\
 & 25\% PV + Bat    & 0.2012 & 14.866726493271 & 7.49623545516769  & 107.2 \\
\hline
\end{tabular}%
}
\end{table}

\begin{figure}
\centering
\includegraphics[width=0.8\textwidth]{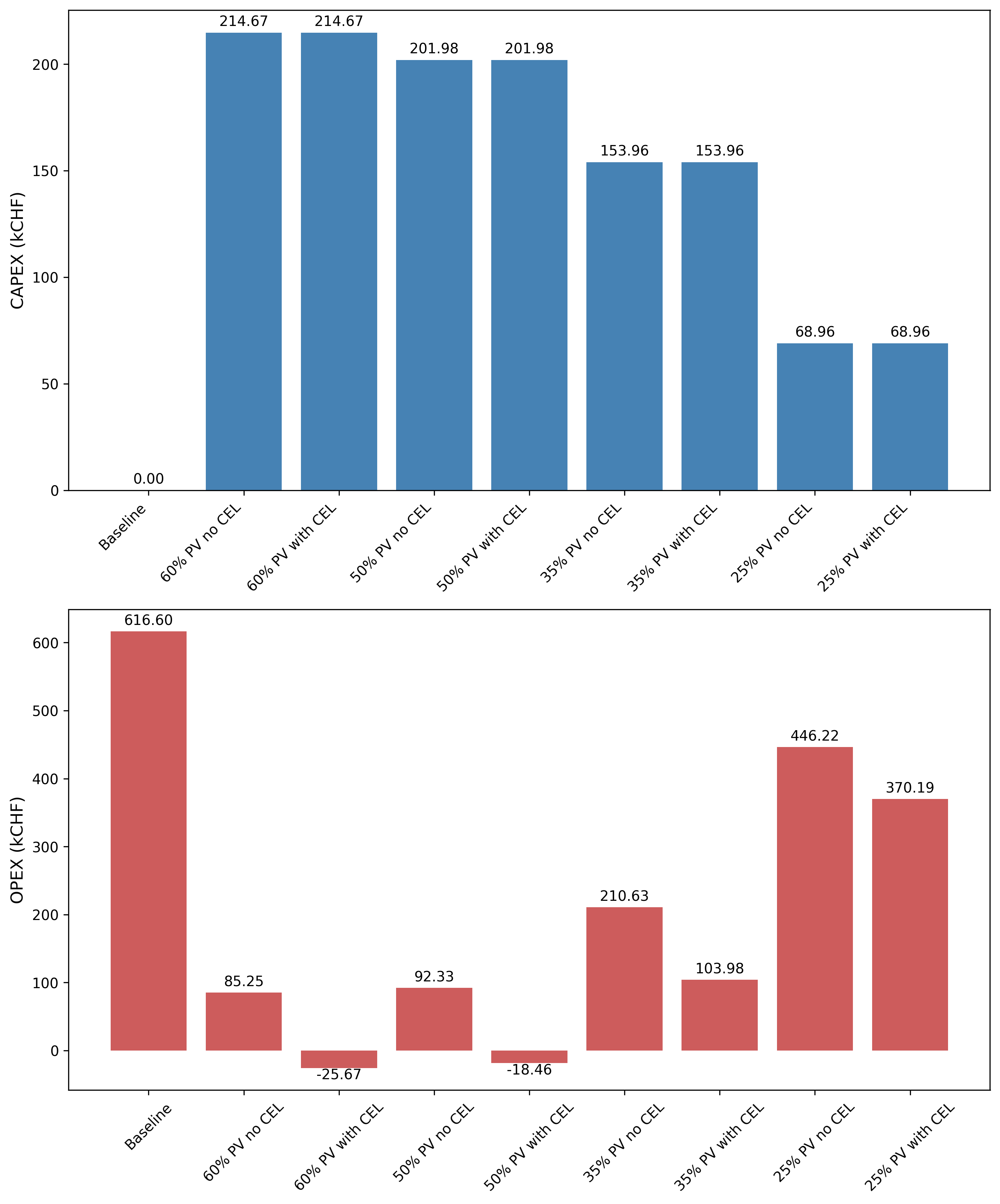}
\caption{CAPEX and OPEX for CEL30}
\label{fig:capex_opex_cel30_partial}
\end{figure}
\begin{figure}
\centering
\includegraphics[width=0.8\textwidth]{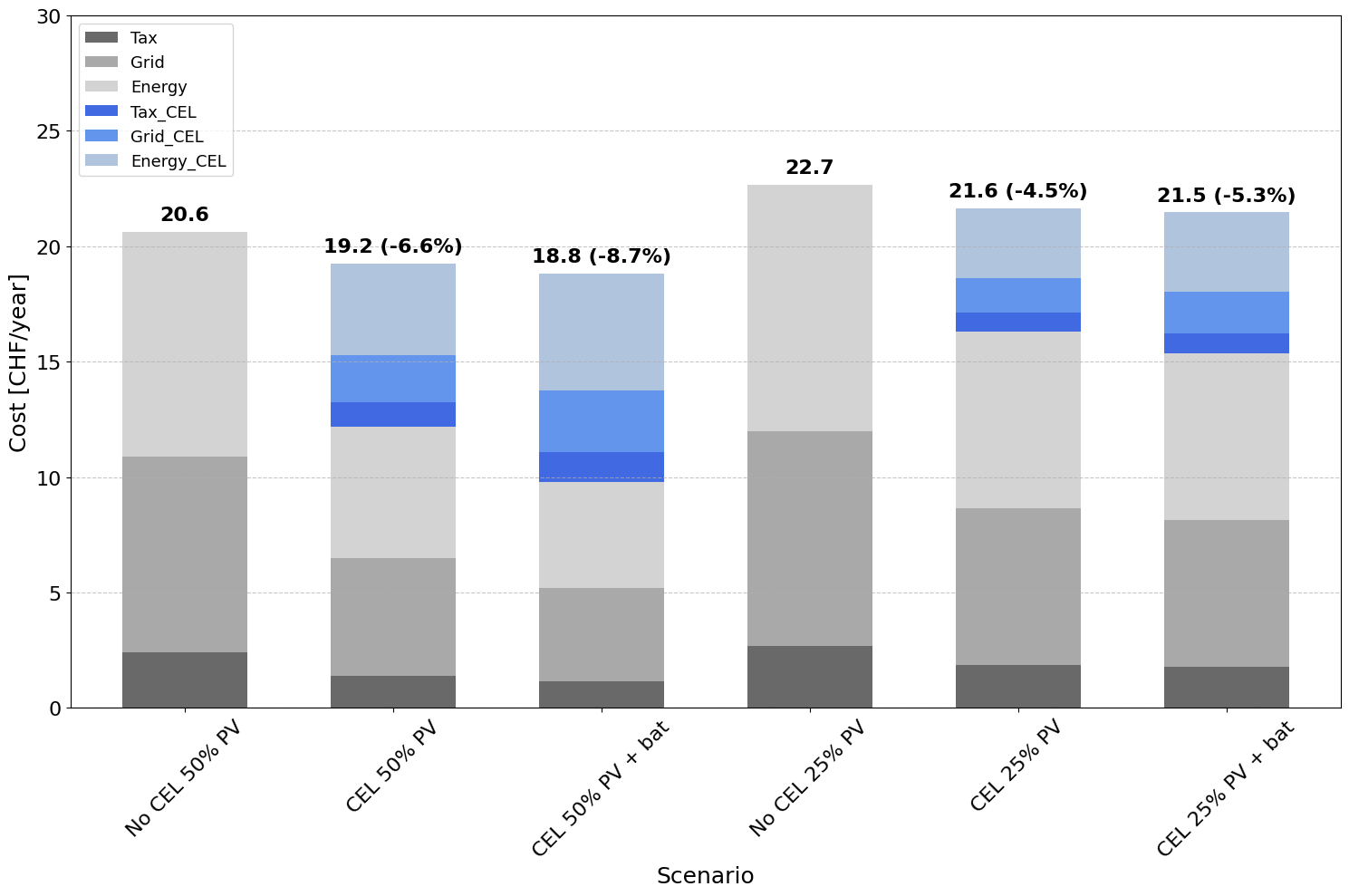}
\caption{Yearly electricity bill for CEL30}
\label{fig:bill_cel30}
\end{figure}

\begin{figure}
\centering
\includegraphics[width=0.8\textwidth]{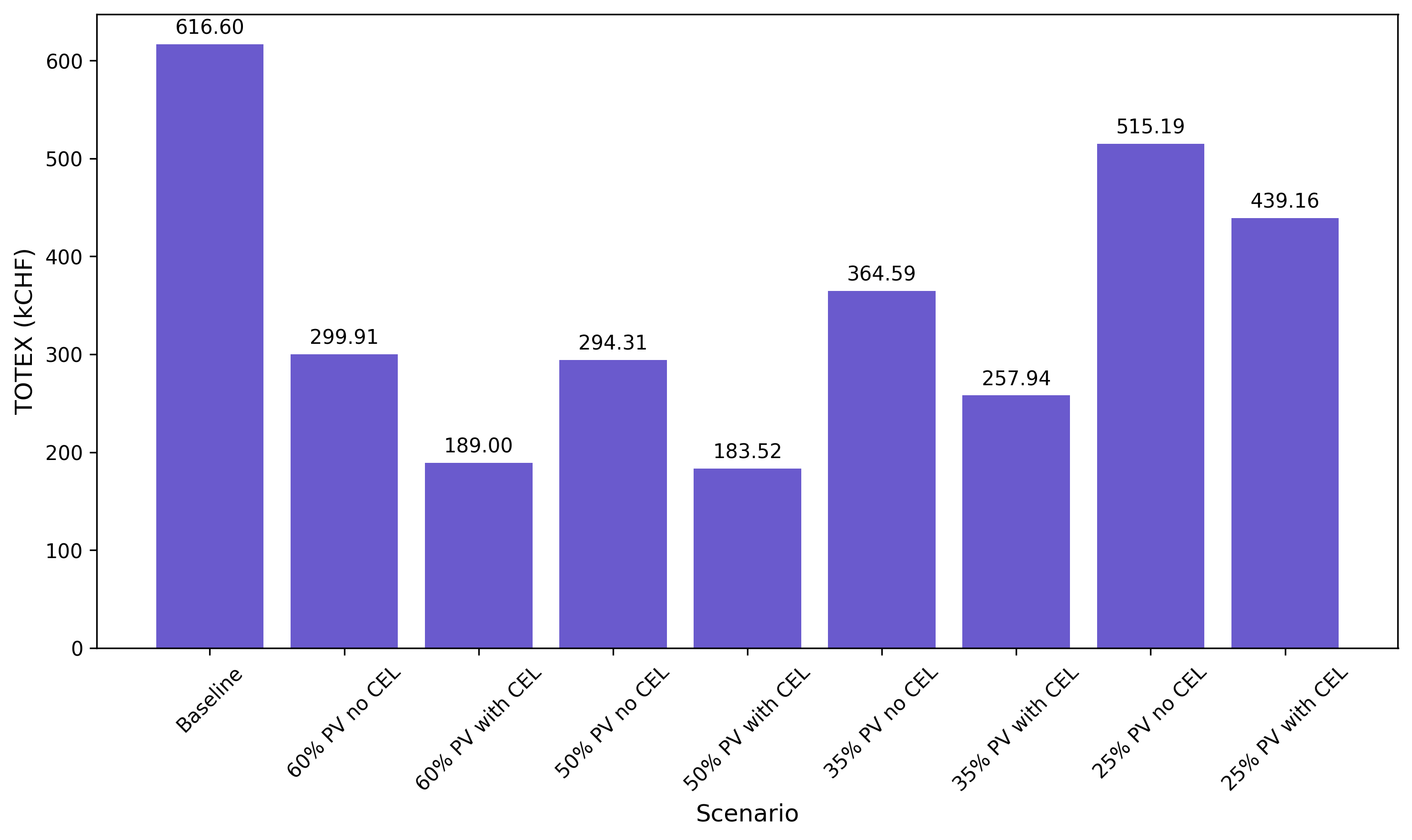}
\caption{TOTEX for CEL30}
\label{fig:totex_cel30}
\end{figure}

\begin{figure}[H]
    \centering
    \includegraphics[width=1\textwidth]{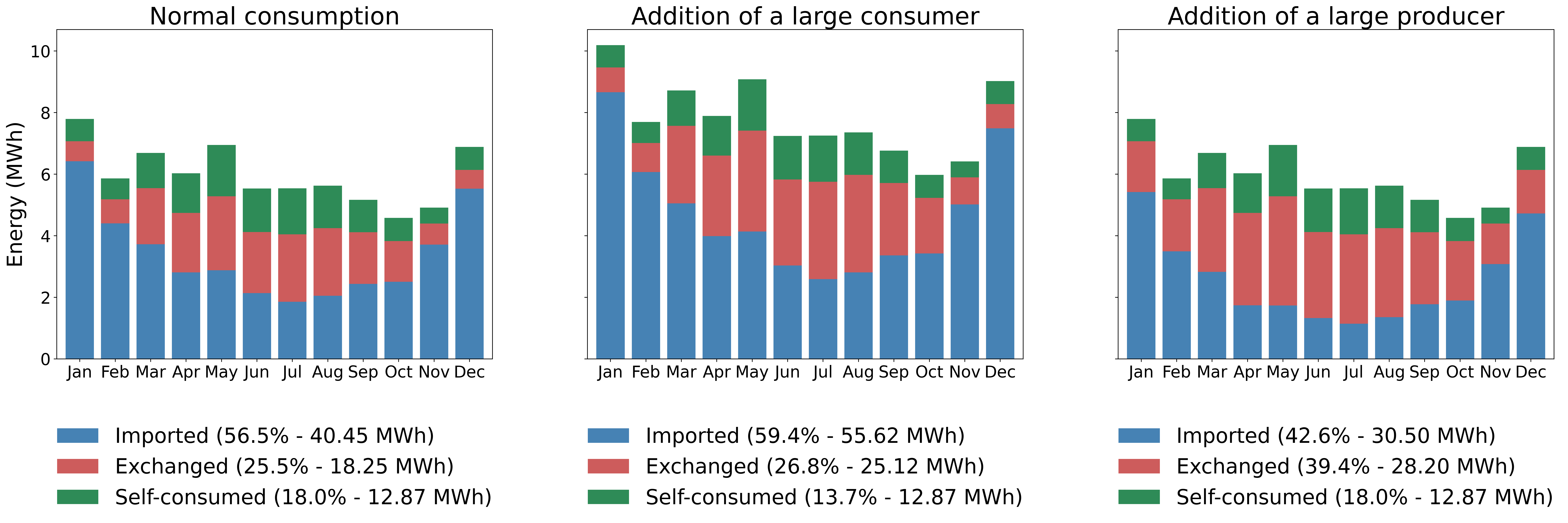}
    \caption{Monthly repartition of imported, exchanged, and self-consumed energy in CEL30 at 50\% PV penetration, comparing the baseline community with the addition of a large producer or a large consumer.}
    \label{fig:cel30_50_energy}
\end{figure}

\begin{figure}[H]
    \centering
    \includegraphics[width=1\textwidth]{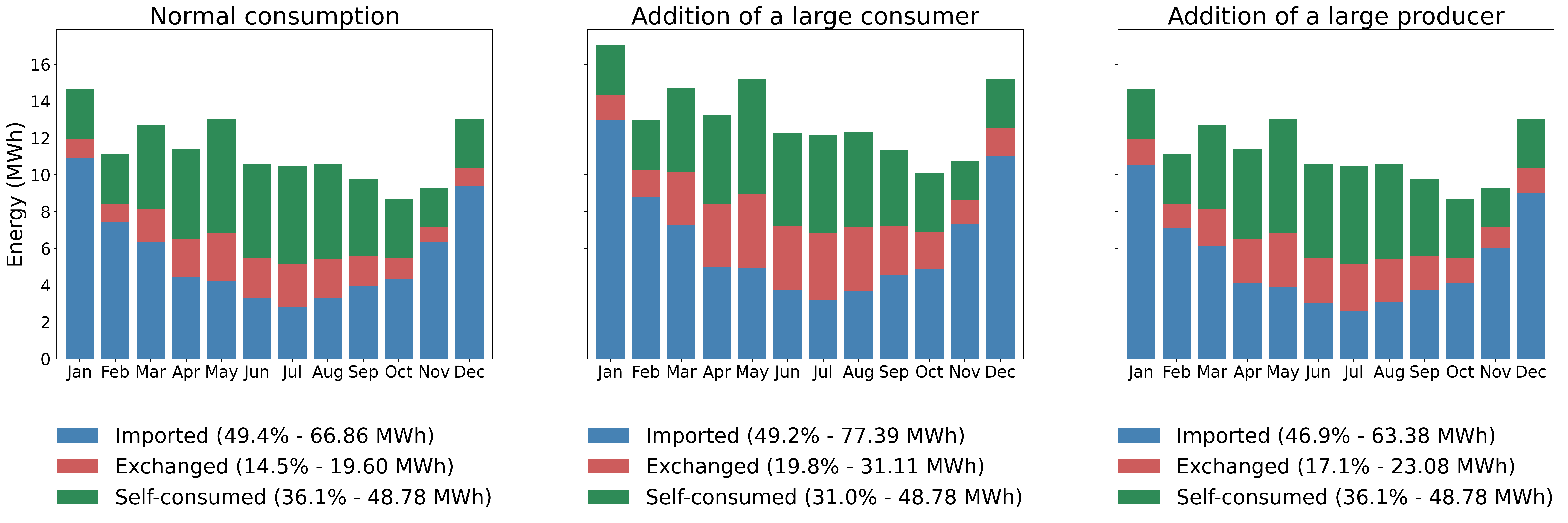}
    \caption{Monthly repartition of imported, exchanged, and self-consumed energy in CEL60 at 50\% PV penetration, comparing the baseline community with the addition of a large producer or a large consumer.}
    \label{fig:cel60_50_energy}
\end{figure}

\begin{table}[H]
\centering
\caption{Techno-economic results for different CEL cases (Dynamic tariff)}
\resizebox{\linewidth}{!}{%
\begin{tabular}{ll
S[round-mode=places,round-precision=4]
S[round-mode=places,round-precision=1]
S[round-mode=places,round-precision=1]
S[round-mode=places,round-precision=1]}
\hline
\textbf{Case} & \textbf{Scenario} & \textbf{LCOE (CHF/kWh)} & \textbf{IRR (\%)} & \textbf{Profit (kCHF)} & \textbf{Bill (kCHF/year)} \\
\hline
\multirow{6}{*}{CEL60}
 & Baseline 50\% PV        & 0.2206 & 7.227  & 274.249 & +2.2 \\
 & 50\% PV with CEL        & 0.1017   & 10.927 & 375.790 & 26.246 \\
 & 50\% PV CEL + Bat       & 0.0941    & 11.049 & 393.250 & 25.214 \\
 & Baseline 25\% PV        & 0.1894   & 11.653 & 93.037  & +4.9 \\
 & 25\% PV with CEL        & 0.1868   & 13.232 & 182.560 & 30.113 \\
 & 25\% PV CEL + Bat       & 0.1824   & 13.314 & 192.685 & 29.298 \\
\hline
\multirow{6}{*}{CEL30}
 & Baseline 50\% PV        & 0.1221  & 11.520 & 154.669 & +6.6 \\
 & 50\% PV with CEL        & 0.1198   & 12.593 & 239.763 & 18.818 \\
 & 50\% PV CEL + Bat       & 0.1134   & 12.703 & 250.157 & 18.214 \\
 & Baseline 25\% PV        & 0.2339 & 12.478 & 45.915  & +0.4 \\
 & 25\% PV with CEL        & 0.2038   & 14.826 & 103.01  & 21.136 \\
 & 25\% PV CEL + Bat       & 0.2011   & 14.859 & 107.429  & 20.7536 \\
\hline
\end{tabular}%
}
\label{tab:lcoe_cel3060_dynamic}
\end{table}
\end{document}